\documentclass[a4paper,reqno,12pt]{amsart}

\usepackage{amssymb,cite,eucal,psfrag,array,setspace}
\usepackage[dvips]{graphicx}

\numberwithin{equation}{section}

\newcommand{\WZW}{Wess-Zumino-Witten }

\newcommand{\group}[1]{\mathsf{#1}}
\newcommand{\alg}[1]{\mathfrak{#1}}

\newcommand{\func}[2]{#1 \left( #2 \right)}

\newcommand{\brac}[1]{\left( #1 \right)}
\newcommand{\sqbrac}[1]{\left[ #1 \right]}

\newcommand{\set}[1]{\left\{ #1 \right\}}

\newcommand{\abs}[1]{\left| #1 \right|}
\newcommand{\ideal}[1]{\left< #1 \right>}
\newcommand{\norm}[1]{\lVert #1 \rVert}
\newcommand{\bilin}[2]{\left( #1 , #2 \right)}

\newcommand{\Det}[1]{\left| #1 \right|}

\newcommand{\conj}[1]{#1^*}

\newcommand{\ZZ}{\mathbb{Z}}

\newcommand{\RR}{\mathbb{R}}
\newcommand{\CC}{\mathbb{C}}

\newcommand{\dd}{\mathrm{d}}
\newcommand{\ii}{\mathfrak{i}}

\newcommand{\eps}{\varepsilon}

\newcommand{\Cox}{\text{h}}
\newcommand{\dCox}{\Cox^{\vee}}

\newcommand{\affine}[1]{\widehat{#1}}

\newcommand{\tenslu}[3]{#1_{#2}^{\phantom{#2} #3} \,}
\newcommand{\tensorcoeff}[3]{\tenslu{N}{#1 #2}{#3}}
\newcommand{\fusioncoeff}[3]{\tenslu{\mathcal{N} \,}{#1 #2}{#3}}
\newcommand{\fusionmat}[1]{\mathcal{N \,}_{#1}}
\newcommand{\fusionring}[1]{\mathcal{F}_{#1}^{\ZZ}}
\newcommand{\fusionideal}[2]{\mathcal{I}_{#1}^{#2}}
\newcommand{\fusionalg}[1]{\mathcal{F}_{#1}^{\, \CC}}
\newcommand{\polyring}[2]{#1 \left[ #2 \right]}

\newcommand{\parD}[2]{\frac{\partial #1}{\partial #2}}
\newcommand{\parDD}[3]{\frac{\partial^2 #1}{\partial #2 \partial #3}}

\newcommand{\LT}[1]{\text{\textsc{lt}} \left( #1 \right)}

\newcommand{\transpose}[1]{#1^{\mathsf{T}}}

\newcommand{\eqnref}[1]{Eqn.~(\ref{#1})}
\newcommand{\eqnDref}[2]{Eqns.~(\ref{#1}) and (\ref{#2})}
\newcommand{\appref}[1]{Appendix~\ref{#1}}
\newcommand{\secref}[1]{Section~\ref{#1}}
\newcommand{\figref}[1]{Figure~\ref{#1}}

\newcommand{\propref}[1]{Proposition~\ref{#1}}

\DeclareMathOperator{\rank}{rank}

\newtheorem{theorem}{Theorem}

\newtheorem{proposition}[theorem]{Proposition}

\begin{document}

\title{Presentations of Wess-Zumino-Witten Fusion Rings}

\author[P Bouwknegt]{Peter Bouwknegt}

\address[Peter Bouwknegt]{Department of Theoretical Physics, Research School of Physical Sciences and Engineering, and Department of Mathematics, Mathematical Sciences Institute \\
Australian National University \\
Canberra, ACT 0200 \\
Australia}

\email{peter.bouwknegt@anu.edu.au}

\author[D Ridout]{David Ridout}

\address[David Ridout]{Department of Physics and Mathematical Physics \\
University of Adelaide \\
Adelaide, SA 5005 \\
Australia,
and Department of Mathematics \\
La Trobe University \\
Bundoora, Melbourne, Vic 3086 \\
Australia}

\curraddr{D\'{e}partement de Physique \\ Universit\'{e} Laval \\ Qu\'{e}bec, G1S 7P4 \\ Canada}

\email{darid@phy.ulaval.ca}

\begin{abstract}
The fusion rings of Wess-Zumino-Witten models are re-examined.  Attention is drawn to the difference between fusion rings over $\ZZ$ (which are often of greater importance in applications) and fusion algebras over $\CC$.  Complete proofs are given characterising the fusion algebras (over $\CC$) of the $\func{\group{SU}}{r+1}$ and $\func{\group{Sp}}{2r}$ models in terms of the fusion potentials, and it is shown that the analagous potentials cannot describe the fusion algebras of the other models.  This explains why no other representation-theoretic fusion potentials have been found.

Instead, explicit generators are then constructed for general WZW fusion rings (over $\ZZ$).  The Jacobi-Trudy identity and its $\func{\group{Sp}}{2r}$ analogue are used to \emph{derive} the known fusion potentials.  This formalism is then extended to the WZW models over the spin groups of odd rank, and explicit presentations of the corresponding fusion rings are given.  The analogues of the Jacobi-Trudy identity for the spinor representations (for all ranks) are derived for this purpose, and may be of independent interest.
\end{abstract}

\maketitle

\onehalfspacing

\section{Introduction} \label{secIntro}

The fusion process is a fundamental ingredient in the standard description of all rational conformal field theories.  Roughly speaking, the fusion coefficient $\fusioncoeff{a}{b}{c}$ counts the multiplicity with which the family of fields $\phi_{c}$ appears in the operator product expansion of a field from family $\phi_{a}$ with a field from family $\phi_{b}$.  This is succinctly written as a fusion rule:
\begin{equation}\label{eqAa-PB}
\phi_{a} \times \phi_{b} = \sum_c \fusioncoeff{a}{b}{c} \phi_{c}.
\end{equation}
This definition makes clear the fact that fusion coefficients are non-negative integers.  Of course, one can define fusion in a more mathematically precise manner in terms of the Grothendieck ring of a certain abelian braided monoidal category that appears in the vertex operator algebra formulation of conformal field theory.  However, we will not have need for such sophistication in what follows.  For our purposes a fusion ring is defined by \eqnref{eqAa-PB}, where the coefficients $\fusioncoeff{a}{b}{c}$ are explicitly given.

The standard assumptions and properties of the operator product expansion then translate into properties of the fusion coefficients.  It is convenient to express these in terms of matrices $\fusionmat{a}$ defined by $\sqbrac{\fusionmat{a}}_{bc} = \fusioncoeff{a}{b}{c}$.  We assume that the identity field is in the theory; the corresponding family is denoted by $\phi_{0}$, and $\fusionmat{0}$ is therefore the identity matrix.  Commutativity and associativity of the operator product expansion translate into
\begin{equation*}
\fusionmat{a} \fusionmat{b} = \fusionmat{b} \fusionmat{a} \qquad \text{and} \qquad \fusionmat{a} \fusionmat{b} = \sum_c \fusioncoeff{a}{b}{c} \fusionmat{c},
\end{equation*}
respectively.  Last, given a family $\phi_{a}$, there is a unique family $\phi_{a^+}$ such that their operator product expansions contain fields from the family $\phi_{0}$ with multiplicity one (this is effectively just the normalisation of the two-point function).  It follows\footnote{Here we are implicitly excluding logarithmic conformal field theories from our considerations.} that $\fusionmat{a^+} = \transpose{\fusionmat{a}}$, where $\transpose{}$ denotes transposition.

These matrices thus form a commuting set of normal matrices, and so may be simultaneously diagonalised by a unitary matrix $U$.  The diagonalisation $\fusionmat{a} U = U D_a$ ($D_a$ diagonal) is equivalent to $\sum_c \fusioncoeff{a}{b}{c} U_{cd} = U_{bd} \lambda^{\brac{a}}_d$ where $\lambda^{\brac{a}}_d$ are the eigenvalues of $\fusionmat{a}$.  Putting $b = 0$ then gives $U_{ad} = U_{0 d} \lambda^{\brac{a}}_d$, which determines the eigenvalues completely (if $U_{0 d}$ were to vanish, $U_{ad}$ would vanish for all $a$ contradicting unitarity).

The celebrated Verlinde conjecture \cite{VerFus} identifies the diagonalising matrix $U$ with the S-matrix $S$ describing the transformations of the characters of the chiral symmetry algebra induced by the modular transformation $\tau \mapsto -1 / \tau$.  This gives a closed expression for the fusion coefficients:
\begin{equation} \label{eqnVerlinde}
\fusioncoeff{a}{b}{c} = \sum_d \frac{S_{ad} S_{bd} \conj{S_{cd}}}{S_{0d}}.
\end{equation}
It is worthwhile noting that the Verlinde conjecture has recently been proved for a fairly wide class of conformal field theories (in the vertex operator algebra approach) \cite{HuaVer}.

Mathematically, these families with their fusion product define a finitely-generated, associative, commutative, unital ring.  Moreover, this \emph{fusion ring} is freely generated as a $\ZZ$-module (abelian group), and possesses a distinguished ``basis'' in which the structure constants are all non-negative integers.  The matrices $\fusionmat{a}$ introduced above correspond to this basis in the regular representation of the fusion ring.  It is often convenient to generalise this structure to a \emph{fusion algebra} (also known as a \emph{Verlinde algebra}) by allowing coefficients in an algebraically closed field, $\CC$ say.  We will denote the fusion ring by $\fusionring{}$, and the corresponding fusion algebra (over $\CC$) by $\fusionalg{} = \fusionring{} \otimes_{\ZZ} \CC$.  It is important to note that the structure which arises naturally in applications is the fusion ring, and that the fusion algebra is just a useful mathematical construct.

One of the first advantages in considering $\fusionalg{}$ is that it contains the elements \cite{FucFus}
\begin{equation}\label{eqnFusIdempotents}
\pi_a = S_{0 a} \sum_b \conj{S_{ab}} \phi_{b},
\end{equation}
where the sum is over the distinguished basis of $\fusionalg{}$.  A quick calculation shows that the $\pi_a$ then form a basis of orthogonal idempotents:  $\pi_a \times \pi_b = \delta_{ab} \pi_b$.  It follows that there are no non-zero nilpotent elements in $\fusionalg{}$, and hence the same is true for $\fusionring{}$.

Since the fusion algebra is finitely-generated, associative, and commutative, it
may be presented as a free polynomial ring (over $\CC$) in its generators,
modulo an ideal $\fusionideal{}{\CC}$.  The lack of non-trivial nilpotent
elements implies that this ideal has the property that whenever some positive power of a polynomial belongs to the ideal, so does the polynomial itself.  That is, the ideal is \emph{radical}, hence completely determined by the variety of points (in $\CC^n$) at which every polynomial in the ideal vanishes \cite{CoxIde}.  This variety will be referred to as the \emph{fusion variety}.

As $\fusionalg{}$ is a finite-dimensional vector space over $\CC$, it follows that the fusion variety consists of a finite number of points, one for each basis element \cite{CoxIde}.  Since the $\pi_a$ of \eqnref{eqnFusIdempotents} form a basis of idempotents, they correspond to polynomials which take the values $0$ and $1$ on the fusion variety.  Their \emph{supports} (points of the fusion variety where the representing polynomials take value $1$) cannot be empty, and their orthogonality ensures that their supports must be disjoint.  This forces the supports to consist of a single point, different for each $\pi_a$.  We denote this point of the fusion variety by $v^a$.  It now follows from inverting \eqnref{eqnFusIdempotents} that the polynomial $p_a$ representing $\phi_{a}$ takes the value $\lambda_b^{\brac{a}} = S_{ab} / S_{0 b}$ at $v^b$.

Suppose now that there is a subset $\set{\phi_{a_i} \colon i = 1 , \ldots , r}$ of the $\phi_{a}$ which generates the entire fusion algebra.  If we take the free polynomial ring to be $\polyring{\CC}{\phi_{a_1} , \ldots , \phi_{a_r}}$, then the coordinates of the fusion variety are just
\begin{equation*}
v^b_i = \func{p_{a_i}}{v^b} = \frac{S_{a_i b}}{S_{0 b}}.
\end{equation*}
This proves the following result of Gepner \cite{GepFus}:
\begin{proposition} \label{propGepner1}
$\fusionalg{} \cong \polyring{\CC}{\phi_{a_1} , \ldots , \phi_{a_r}} / \fusionideal{}{\CC}$, where $\fusionideal{}{\CC}$ is the (radical) ideal of polynomials vanishing on the points
\begin{equation*}
\set{\brac{\frac{S_{a_1 b}}{S_{0 b}} , \ldots , \frac{S_{a_r b}}{S_{0 b}}} \in \CC^r}.
\end{equation*}
\end{proposition}

Notice that this result only characterises the fusion algebra.  The fusion ring may likewise be represented as a quotient of $\polyring{\ZZ}{\phi_{a_1} , \ldots , \phi_{a_r}}$, where the fusion ideal is given by $\fusionideal{}{\ZZ} = \fusionideal{}{\CC} \cap \polyring{\ZZ}{\phi_{a_1} , \ldots , \phi_{a_r}}$ \cite{FucFus}.  The fusion ideal over $\ZZ$ thus inherits the property from $\fusionideal{}{\CC}$ that if any integral multiple of a polynomial is in the ideal, then so is the polynomial itself.  This ensures that the quotient is a free $\ZZ$-module, as required.  By analogy with radical ideals (and for wont of a better name), we will refer to ideals with this property as being \emph{dividing}.

In this paper, we are interested in the fusion rings of \WZW models.  These are conformal field theories defined on a group manifold $\group{G}$ (which we will take to be simply-connected, connected, and compact), and parametrised by a positive integer $k$ called the level.  Our motivation derives from the determination of the dynamical charge group of a certain class of D-brane in these theories.  The brane charges \cite{PolDir,MinKTh} can be computed explicitly, and the order of the charge group can be shown to be constrained by the fusion rules \cite{BouNot,FreBra}.  A suitably detailed understanding of the structure of the fusion rules therefore makes the computation of the charge group possible.  This was achieved for the models based on the groups $\group{G} = \func{\group{SU}}{r+1}$ in \cite{FreBra}, and the general case in \cite{BouDBr02}.

However, the general charge group computations have only been rigorously proved for $\group{G} = \func{\group{SU}}{r+1}$ and\footnote{In this paper we denote by $\func{\group{Sp}}{2r}$ the (unique up to isomorphism) connected, simply-connected, compact Lie group whose Lie algebra is $\func{\alg{sp}}{2r}$.} $\func{\group{Sp}}{2r}$, essentially because the detailed structure of the fusion rules associated with the other groups is not well understood.  The aim of this paper is to re-examine the cases which have been described, and try to elucidate a corresponding detailed structure in other cases.

The field families of a level-$k$ \WZW model on the group manifold $\group{G}$ are conveniently labelled by an integrable highest weight representation of the associated untwisted affine Lie algebra $\affine{\alg{g}}$, hence by the projection of the highest weight onto the weight space of the horizontal subalgebra $\alg{g}$ (which will be identified with the Lie algebra of $\group{G}$).  In other words, the abstract elements naturally appearing in the fusion rules may be identified with the integral weights (of $\alg{g}$) in the closed fundamental affine alcove.  We denote this set of weights by $\affine{\group{P}}_k$.  In what follows, it will usually prove more useful to regard these weights as the integral weights in the \emph{open, shifted} fundamental alcove.  Concretely,
\begin{equation*}
\affine{\group{P}}_k = \set{\lambda \in \group{P} \colon \bilin{\lambda + \rho}{\alpha_i} > 0 \text{ for all } i, \text{ and } \bilin{\lambda + \rho}{\theta} < k + \dCox},
\end{equation*}
where $\group{P}$ is the weight lattice, $\alpha_i$ are the simple roots, $\theta$ denotes the highest root, $\rho$ the Weyl vector, and $\dCox$ is the dual Coxeter number of $\alg{g}$.  The inner product on the weight space is normalised so that $\bilin{\theta}{\theta} = 2$.

For these \WZW models, the Verlinde conjecture was proven in \cite{TsuCon,FalPro,BeaCon}.  By combining this with the Kac-Peterson formula \cite{KacPetInf} for the \WZW S-matrix elements,
\begin{equation} \label{eqnKacPeterson}
S_{\lambda \mu} = \func{C}{\affine{\alg{g}} , k} \sum_{w \in \group{W}} \det w \ e^{-2 \pi \ii \bilin{\func{w}{\lambda + \rho}}{\mu + \rho} / \brac{k + \dCox}}
\end{equation}
(here $\func{C}{\affine{\alg{g}} , k}$ is a constant and $\group{W}$ is the Weyl group of $\group{G}$), one can derive a very useful expression for the fusion coefficients, known as the Kac-Walton formula \cite{KacInf,WalFus,WalAlg,FucWZW,FurQua}:
\begin{equation} \label{eqnKacWalton}
\fusioncoeff{\lambda}{\mu}{\nu} = \sum_{\affine{w} \in \affine{\group{W}}_k} \det \affine{w} \ \tensorcoeff{\lambda}{\mu}{\affine{w} \cdot \nu}.
\end{equation}
This formula relates the fusion coefficients to the tensor product multiplicities $\tensorcoeff{\lambda}{\mu}{\nu}$ of the irreducible representations of the group $\group{G}$ (or its Lie algebra $\alg{g}$), via the shifted action of the affine Weyl group $\affine{\group{W}}_k$ at level $k$, $\affine{w} \cdot \nu = \func{\affine{w}}{\nu + \rho} - \rho$.

The Kac-Walton formula suggests that for \WZW models, it may be advantageous to choose the free polynomial ring appearing in \propref{propGepner1} to be the complexified representation ring (character ring) of $\group{G}$.  The character of the irreducible representation of highest weight $\lambda$ is given by
\begin{equation*}
\chi_{\lambda} = \sum_{\mu \in P_{\lambda}} e^{\mu} = \frac{\sum_{w \in \group{W}} \det w \ e^{\func{w}{\lambda + \rho}}}{\sum_{w \in \group{W}} \det w \ e^{\func{w}{\rho}}},
\end{equation*}
where $P_{\lambda}$ is the set of weights of the representation with multiplicity (and the second equality is the Weyl character formula).  The character ring is freely generated by the characters $\chi_{\Lambda_i} \equiv \chi_i$ ($i = 1 , \ldots , r = \rank \group{G}$) of the representations whose highest weights are the fundamental weights $\Lambda_i$ of $\group{G}$.  Gepner's result for \WZW models may therefore be recast in the form:
\begin{proposition} \label{propGepner}
The fusion algebra of a level-$k$ \WZW model is given by $\fusionalg{k} \cong \polyring{\CC}{\chi_1 , \ldots , \chi_r} / \fusionideal{k}{\CC}$, where $\fusionideal{k}{\CC}$ is the (radical) ideal of polynomials vanishing on the points
\begin{equation*}
\set{\brac{\frac{S_{\Lambda_1 \lambda}}{S_{0 \lambda}} , \ldots , \frac{S_{\Lambda_r \lambda}}{S_{0 \lambda}}} \in \CC^r \colon \lambda \in \affine{\group{P}}_k}.
\end{equation*}
\end{proposition}
We will likewise denote a level-$k$ \WZW fusion ring by $\fusionring{k}$ and the corresponding fusion ideal of $\polyring{\ZZ}{\chi_1 , \ldots , \chi_r}$ by $\fusionideal{k}{\ZZ}$.

We are interested in explicit sets of generators for these fusion ideals (over $\CC$ and $\ZZ$).  Given a candidate set of elements in $\fusionideal{k}{\CC}$, the verification that this set is generating may be broken down into three parts:  First, one checks that each element vanishes on the fusion variety.  Second, one must show that these elements do not collectively vanish anywhere else.  Third, the ideal generated by this candidate set must be verified to be radical.  This last step is always necessary because there is generically an infinite number of ideals corresponding to a given variety (consider the ideals $\ideal{x^n} \subset \polyring{\CC}{x}$ which all vanish precisely at the origin).  It should be clear that verifying radicality does not consist of the trivial task of checking that the candidate generating set contains no powers of polynomials (consider $\ideal{x^2 + y^2 , 2 x y} \subset \polyring{\CC}{x,y}$).

For the $\func{\group{SU}}{r+1}$ and $\func{\group{Sp}}{2r}$ fusion algebras, generating sets for $\fusionideal{k}{\CC}$ have been postulated in \cite{GepFus,BouTop,GepSym} as the partial derivatives of a \emph{fusion potential}.  The first step of the verification process is well-documented there, the second step appears somewhat sketchy, and the third does not seem to have appeared in the literature at all.  We rectify this in \secref{secFusAlg}.  The methods we employ are then used to show why analogous potentials have not been found for the other groups, despite several attempts \cite{CreFus,MlaInt}.

However, we would like to repeat our claim that it is the fusion ring which is of physical interest in applications, and the above verification process does not allow us to conclude that a set of elements is generating over $\ZZ$.  In other words, a set of generators for $\fusionideal{k}{\CC}$ need not form a generating set for $\fusionideal{k}{\ZZ}$, even if the set consists of polynomials with integral coefficients (a simple example would be if $\fusionideal{k}{\CC} = \ideal{x + y , x - y} \subset \polyring{\CC}{x,y}$ then $\fusionideal{k}{\ZZ} \neq \ideal{x + y , x - y} \subset \polyring{\ZZ}{x,y}$ as this latter ideal is not dividing).  This consideration also seems to have been overlooked in the literature, and is, in our opinion, quite a serious omission.  We will rectify this situation in \secref{secFusRing} by removing the need to postulate a candidate set of generators; instead, we shall derive generating sets \emph{ab initio}.

In the cases $\group{G} = \func{\group{SU}}{r+1}$ and $\func{\group{Sp}}{2r}$, some simple manipulations will allow us to reduce the number of generators in these sets drastically.  We will see that these manipulations reproduce the aforementioned fusion potentials.  Our results therefore constitute the first complete derivation of this description from first principles, and we emphasise that this derivation holds over $\ZZ$.  The results to this point have already been detailed in \cite{RidPhD}.  We then detail the analagous manipulations for $\func{\group{Spin}}{2r+1}$ in \secref{secFusSpin}, producing a relatively small set of explicit generators for the corresponding fusion ideal.  It is not clear to us whether these generators are related to a description by fusion potentials.  The manipulations essentially rely upon the application of a class of identities generalising the classical Jacobi-Trudy identity (which we will collectively refer to as Jacobi-Trudy identities).  Many of these are well-known \cite{WeyCla}, but we were unable to find identities for spinor representations in the literature, so we include derivations in \appref{secJT}.  We also include the corresponding identities for $\func{\group{Spin}}{2r}$, as they may be of independent interest.

\section{Presentations of Fusion Algebras} \label{secFusAlg}

In this section, we consider the description of the fusion ideals $\fusionideal{k}{\CC}$ by fusion potentials.  We introduce the potentials for the \WZW models over the groups $\func{\group{SU}}{r+1}$ and $\func{\group{Sp}}{2r}$, and verify that the induced ideals vanish precisely on the fusion variety, \emph{and} are radical.  We then investigate the obvious class of analogous potentials for \WZW models over other groups, and show that in these cases, no potential in this class correctly describes the fusion algebra.  Readers that are only interested in fusion \emph{rings} and presentations of the ideals $\fusionideal{k}{\ZZ}$ should skip to \secref{secFusRing}.

\subsection{Fusion Potentials} \label{secFusPot}

For \WZW models over $\func{\group{SU}}{r+1}$ and $\func{\group{Sp}}{2r}$, the fusion ideal is supposed to be generated by the partial derivatives (with respect to the characters $\chi_i$ of the fundamental representations) of a single polynomial, called the \emph{fusion potential}.  At level $k$, \cite{GepFus} gives the $\func{\group{SU}}{r+1}$-potential as
\begin{equation} \label{eqnFusPotSU}
\func{V_{k+r+1}}{\chi_1 , \ldots , \chi_r} = \frac{1}{k+r+1} \sum_{i=1}^{r+1} q_i^{k+r+1},
\end{equation}
where the $q_i$ are the (formal) exponentials of the weights $\eps_i$ of the defining representation (whose character is $\chi_1$).  Note that $q_1 \cdots q_{r+1} = 1$.  The $\eps_i$ are permuted by the Weyl group $\group{W} = \group{S}_{r+1}$ of $\func{\group{SU}}{r+1}$, and $\group{W}$ acts analogously on the $q_i$.  Therefore, $V_{k+r+1}$ is clearly $\group{W}$-invariant, hence is indeed a polynomial in the $\chi_i$ \cite{BouLie2}.

The level-$k$ $\func{\group{Sp}}{2r}$-potential is given in \cite{BouTop,GepSym} as
\begin{equation} \label{eqnFusPotSp}
\func{V_{k+r+1}}{\chi_1 , \ldots , \chi_r} = \frac{1}{k+r+1} \sum_{i=1}^r \sqbrac{q_i^{k+r+1} + q_i^{-\brac{k+r+1}}},
\end{equation}
where the $q_i$ and $q_i^{-1}$ refer to the (formal) exponentials of the weights $\pm \eps_i$ of the defining representation of $\func{\group{Sp}}{2r}$ (whose character is again $\chi_1$).  The Weyl group $\group{W} = \group{S}_r \ltimes \ZZ_2^r$ acts on the $\eps_i$ by permutation ($\group{S}_r$) and negation (each $\ZZ_2$ sends one $\eps_i$ to $-\eps_i$ whilst leaving the others invariant).  We see again that the given potential is a $\group{W}$-invariant, hence a polynomial in the $\chi_i$.

These potentials are obviously best handled with generating functions.  We also note that these potentials may be unified as
\begin{equation} \label{eqnFusPotSUSp}
\func{V_{k+\dCox}}{\chi_1 , \ldots , \chi_r} = \frac{1}{k+\dCox} \sum_{\mu \in P_{\Lambda_1}} e^{\brac{k+\dCox} \mu},
\end{equation}
where $P_{\lambda}$ denotes the set of weights of the irreducible representation of highest weight $\lambda$.  Putting this form into a generating function (and dropping the explicit $\chi_i$ dependence) gives
\begin{equation*}
\func{V}{t} = \sum_{m=1}^{\infty} \brac{-1}^{m-1} V_m t^m = \log \sqbrac{\prod_{\mu \in P_{\Lambda_1}} \brac{1 + e^{\mu} t}}.
\end{equation*}
This generating function may therefore be expressed in terms of the characters of the exterior powers of the defining representation.  These exterior powers are well-known \cite{FulRep}, and give
\begin{align}
\func{\group{SU}}{r+1} & \text{:} & \func{V}{t} &= \log \sqbrac{\sum_{n=0}^{r+1} \chi_n t^n}, \label{eqnFusPotGFSU}
\intertext{where $\chi_0 = \chi_{r+1} = 1$, and}
\func{\group{Sp}}{2r} & \text{:} & \func{V}{t} &= \log \sqbrac{\sum_{n=0}^{r-1} E_n \brac{t^n + t^{2r-n}} + E_r t^r}, \label{eqnFusPotGFSp}
\end{align}
where $\chi_0 = 1$, $\chi_n = 0$ for all $n < 0$, and $E_n = \chi_n + \chi_{n-2} + \chi_{n-4} + \ldots$.

At this point it should be mentioned that there is an explicit construction for arbitrary rational conformal field theories \cite{AhaGen}, which determines a function whose derivatives vanish on the fusion variety.  This construction, however, requires an explicit knowledge of the S-matrix elements, and is quite unwieldy (as compared with the above potentials).  Indeed, it also seems to possess significant ambiguities, and it is not clear how to fix this so as to find a potential with a representation-theoretic interpretation.  In any case, it also appears to be difficult to determine if these ideals thus obtained are radical or dividing, so we will not consider this construction any further.  There is also a paper \cite{CreFus} postulating simple potentials for every \WZW model, similar in form to those of \eqnDref{eqnFusPotSU}{eqnFusPotSp}.  But, as pointed out in \cite{MlaInt}, the partial derivatives of the potentials given do not always vanish on the fusion variety, and so cannot generate the fusion ideal.  In \cite{MlaInt}, fusion potentials are presented for rings related to the fusion rings of the \WZW models over the special orthogonal groups.  Unfortunately, their method fails to give the fusion rings for the special orthogonal groups.  We will see in in \secref{secGenPot} why this is the case.

\subsection{Verification} \label{secFusPotVerify}

Let us first establish that the ideals defined by the potentials given in \eqnDref{eqnFusPotSU}{eqnFusPotSp} vanish on their respective fusion varieties.  From \propref{propGepner}, the points of the fusion variety have coordinates
\begin{equation*}
v^{\lambda}_i = \frac{S_{\Lambda_i \lambda}}{S_{0 \lambda}} = \func{\chi_i}{-2 \pi \ii \frac{\lambda + \rho}{k + \dCox}},
\end{equation*}
where the second equality follows readily from Weyl's character formula and \eqnref{eqnKacPeterson}.  It follows that the fusion potentials should have critical points precisely when the characters are evaluated at $\xi_{\lambda} = -2 \pi \ii \brac{\lambda + \rho} / \brac{k + \dCox}$, for $\lambda \in \affine{\group{P}}_k$.  In fact, the functions $\varkappa_i$ defined by
\begin{equation*}
\func{\varkappa_i}{\lambda} = \func{\chi_{i}}{-2 \pi \ii \frac{\lambda + \rho}{k + \dCox}} = \sum_{\mu \in P_{\Lambda_i}} e^{-2 \pi \ii \bilin{\mu}{\lambda + \rho} / \brac{k + \dCox}}
\end{equation*}
are invariant under the shifted action of the affine Weyl group $\affine{\group{W}}_k$.  Thus, the potentials should have critical points when evaluated at $\chi_i = \func{\varkappa_i}{\lambda}$, for any $\lambda \in \group{P}$ which is \emph{not} on a shifted alcove boundary.

We denote the gradient operations with respect to the fundamental characters $\chi_i$ and the Dynkin labels $\lambda_j$ by $\nabla_{\chi}$ and $\nabla_{\lambda}$ respectively, and the jacobian matrix of the functions $\varkappa_i$ with respect to the $\lambda_j$ by $J$.  From the chain rule, it follows that if the potential has a critical point with respect to $\lambda$ at which $J$ is non-singular, then this is also a critical point with respect to the fundamental characters.  It is therefore necessary to determine when $J$ becomes singular.

Explicit calculation shows that the jacobian, as a function on the weight space, satisfies
\begin{equation} \label{eqnDetJw}
\func{J}{\func{w}{\nu}} = \func{J}{\nu} w,
\end{equation}
hence $\det J$ is anti-invariant under the Weyl group $\group{W}$ (here, $w$ on the right hand side refers to the matrix representation of $w$ with respect to the basis of fundamental weights).  It is therefore a multiple of the primitive anti-invariant element \cite{BouLie2}, and by comparing leading terms, we arrive at
\begin{equation*}
\det J = \brac{\frac{-2 \pi \ii}{k + \dCox}}^r \frac{1}{\abs{\group{P} / \group{Q}^{\vee}}} \prod_{\alpha \in \Delta_+} \brac{e^{\alpha / 2} - e^{-\alpha / 2}},
\end{equation*}
where $\group{Q}^{\vee}$ is the coroot lattice and $\Delta_+$ are the positive roots of $\alg{g}$ (explicit details may be found in \cite{RidPhD}).  Evaluating at $-2 \pi \ii \brac{\lambda + \rho} / \brac{k + \dCox}$, it follows that the jacobian is singular precisely when 
\begin{equation*}
\prod_{\alpha \in \Delta_+} \sin \sqbrac{\pi \frac{\bilin{\alpha}{\lambda + \rho}}{k + \dCox}} = 0.
\end{equation*}
That is, when $\lambda$ is on the boundary of a shifted affine alcove.  Therefore, these boundaries are the only places where a potential may have critical points with respect to $\lambda$ which need not be critical points with respect to the $\chi_i$.

Evaluating the potentials, \eqnref{eqnFusPotSUSp}, as above gives
\begin{equation*}
\func{V_{k + \dCox}}{\func{\varkappa_1}{\lambda} , \ldots , \func{\varkappa_r}{\lambda}} = \frac{1}{k + \dCox} \sum_{\mu \in P_{\Lambda_1}} e^{-2 \pi \ii \bilin{\mu}{\lambda + \rho}} = \frac{1}{k + \dCox} \func{\chi_1}{-2 \pi \ii \brac{\lambda + \rho}}.
\end{equation*}
Note that the level dependence becomes quite trivial.  We now determine the critical points of these potentials with respect to the Dynkin labels $\lambda_j$.
\begin{description}
\item[$\func{\group{Sp}}{2r}$]
The $2r$ weights of the defining representation are the $\eps_j$ and their negatives.  The potentials therefore take the form
\begin{equation*}
\func{V_{k+\dCox}}{-2 \pi \ii \frac{\lambda + \rho}{k + \dCox}} = \frac{2}{k + \dCox} \sum_{j=1}^r \cos \sqbrac{2 \pi \bilin{\eps_j}{\lambda + \rho}}.
\end{equation*}
Critical points therefore occur when
\begin{equation*}
\sum_{j=1}^r \bilin{\Lambda_i}{\eps_j} \sin \sqbrac{2 \pi \bilin{\eps_j}{\lambda + \rho}} = 0,
\end{equation*}
for each $i = 1 , \ldots , r$.  The $\bilin{\Lambda_i}{\eps_j}$ form the entries of a square matrix which is easily seen to be invertible, as $\eps_j = \frac{1}{2} \brac{\alpha_j^{\vee} + \ldots + \alpha_r^{\vee}}$ \cite{BouLie2}.  We therefore have critical points precisely when
\begin{equation*}
\sin \sqbrac{2 \pi \bilin{\eps_j}{\lambda + \rho}} = \sin \sqbrac{\pi \brac{\lambda_j + \rho_j + \ldots + \lambda_r + \rho_r}} = 0,
\end{equation*}
for all $j = 1 , \ldots , r$.  It follows that $\lambda_j + \ldots + \lambda_r \in \ZZ$ for each $j = 1 , \ldots , r$, hence $\lambda \in \group{P}$.
\item[$\func{\group{SU}}{r+1}$]
In this case, the $r+1$ weights of the defining representation are the $\eps_j$, but we have the constraint $\eps_1 + \ldots + \eps_{r+1} = 0$.  Finding the critical points on the weight space is a constrained optimisation problem in $\RR^{r+1}$, so we add a Lagrange multiplier $\Omega$ to the potential:
\begin{equation*}
\func{\widetilde{V}_{k+\dCox}}{-2 \pi \ii \frac{\lambda + \rho}{k + \dCox}} = \frac{1}{k + \dCox} \sum_{j=1}^{r+1} e^{-2 \pi \ii \bilin{\eps_j}{\lambda + \rho}} + \Omega \bilin{\lambda}{\eps_1 + \ldots + \eps_{r+1}}.
\end{equation*}
It is now straightforward to show that the critical points are again $\lambda \in \group{P}$, so we leave this as an exercise for the reader.
\end{description}

So, for both $\func{\group{SU}}{r+1}$ and $\func{\group{Sp}}{2r}$, the critical points with respect to $\lambda$ of the potentials of \eqnref{eqnFusPotSUSp} coincide with the weight lattice $\group{P}$.  Every integral weight which is not on a shifted affine alcove boundary therefore corresponds to a critical point with respect to the fundamental characters (since $J$ is non-singular there).  To conclude that the critical points of the potentials coincide with the points of the corresponding fusion varieties, we therefore need to exclude the possibility that an integral weight on a shifted affine alcove boundary can correspond to a critical point with respect to the fundamental characters.  This follows readily from a study of the determinant of the hessian matrix $H_{\lambda} = \brac{\parDD{V_{k+\dCox}}{\lambda_i}{\lambda_j}}$ of the potentials at these points, whose computation we now turn to.
\begin{description}
\item[$\func{\group{SU}}{r+1}$]
Here (indeed, for any simply-laced group), $\group{P}$ coincides with the dual of the root lattice.  Thus, $\lambda \in \group{P}$ implies that $\bilin{\mu}{\lambda + \rho} = \bilin{\Lambda_1}{\lambda + \rho} \pmod{1}$ for all $\mu \in P_{\Lambda_1}$.  It follows that 
\begin{align*}
\brac{H_{\lambda}}_{ij} &= \frac{-4 \pi^2}{k+\dCox} \sum_{\mu \in P_{\Lambda_1}} \bilin{\mu}{\Lambda_i} \bilin{\mu}{\Lambda_j} e^{-2 \pi \ii \bilin{\mu}{\lambda + \rho}} \\
&= \frac{-4 \pi^2}{k+\dCox} e^{-2 \pi \ii \bilin{\Lambda_1}{\lambda + \rho}} I_{\Lambda_1} \bilin{\Lambda_i}{\Lambda_j},
\end{align*}
where $I_{\Lambda_1}$ is the Dynkin index of the irreducible representation of highest weight $\Lambda_1$.  Thus,
\begin{equation*}
\det H_{\lambda} = \brac{\frac{-4 \pi^2 I_{\Lambda_1}}{k+\dCox}}^r \frac{e^{-2 \pi \ii r \bilin{\Lambda_1}{\lambda + \rho}}}{\abs{\group{P} / \group{Q}^{\vee}}} \neq 0,
\end{equation*}
when $\lambda \in \group{P}$.
\item[$\func{\group{Sp}}{2r}$]
The weights of $P_{\Lambda_1}$ take the form $\pm \eps_{\ell} = \pm \frac{1}{2} \brac{\alpha_{\ell}^{\vee} + \ldots + \alpha_r^{\vee}}$, for $\ell = 1 , 2 , \ldots , r$, so $\bilin{\eps_{\ell}}{\Lambda_i} \bilin{\eps_{\ell}}{\Lambda_j} = \frac{1}{4}$ if $i \geqslant \ell$ and $j \geqslant \ell$, and $0$ otherwise.  Computing the hessian as before gives
\begin{equation*}
\brac{H_{\lambda}}_{ij} = \frac{-2 \pi^2}{k+\dCox} \sum_{\ell = 1}^{\min \set{i , j}} \cos \sqbrac{\pi \brac{\lambda_{\ell} + \ldots + \lambda_r + r - \ell + 1}}.
\end{equation*}
Elementary row operations now suffice to compute
\begin{equation*}
\det H_{\lambda} = \brac{\frac{-2 \pi^2}{k+\dCox}}^r \prod_{\ell = 1}^r \cos \sqbrac{\pi \brac{\lambda_{\ell} + \ldots + \lambda_r + r - \ell + 1}},
\end{equation*}
so again $\det H_{\lambda} \neq 0$ on the weight lattice.
\end{description}

Denote the hessian matrix with respect to the $\chi_i$ of the potentials by $H_{\chi}$.  Then, from
\begin{equation*}
\parDD{V_{k+\dCox}}{\lambda_i}{\lambda_j} = \sum_{s,t} \parD{\chi_s}{\lambda_i} \parDD{V_{k+\dCox}}{\chi_s}{\chi_t} \parD{\chi_t}{\lambda_j} + \sum_{\ell} \parD{V_{k+\dCox}}{\chi_{\ell}} \parDD{\chi_{\ell}}{\lambda_i}{\lambda_j},
\end{equation*}
we see that
\begin{equation*}
H_{\lambda} = \transpose{J} H_{\chi} J \qquad \text{when $\nabla_{\chi} V_{k+\dCox} = 0$.}
\end{equation*}
It follows that at the critical points of the potential with respect to the $\chi_i$, 
\begin{equation} \label{eqnHessians}
\det H_{\lambda} = \brac{\det J}^2 \det H_{\chi}.
\end{equation}
Now, we have just demonstrated that $\det H_{\lambda} \neq 0$ on the weight lattice, but we know that $\det J = 0$ on the shifted affine alcove boundaries.  As $\det H_{\chi}$ is a polynomial (hence finite-valued), this forces the conclusion that any integral weight lying on a shifted affine alcove boundary is \emph{not} a critical point of the potential with respect to the $\chi_i$.  Of course, this is exactly what we wanted to show.

To summarise, we have shown that the ideal generated by the derivatives of the potentials given in \eqnDref{eqnFusPotSU}{eqnFusPotSp} vanishes precisely on the fusion variety.  To complete the proof (over $\CC$) that these potentials describe the fusion ideal $\fusionideal{k}{\CC}$, we need to show that this ideal is radical.  Happily, this follows immediately from \eqnref{eqnHessians} and some standard multiplicity theory, specifically the theory of \emph{Milnor numbers} \cite{CoxUsi,MilSin}:  The ideal generated by the derivatives of a potential is radical if and only if the hessian of the potential is non-singular at each point of the corresponding (zero-dimensional) variety.  Since $H_{\lambda}$ and $J$ are non-singular at the points of the fusion variety, $H_{\chi}$ is non-singular there by \eqnref{eqnHessians}, and we are done.  The ideals are radical, so the potentials given by \eqnDref{eqnFusPotSU}{eqnFusPotSp} correctly describe the fusion algebras of $\func{\group{SU}}{r+1}$ and $\func{\group{Sp}}{2r}$ (respectively).

\subsection{A Class of Candidate Potentials} \label{secGenPot}

In searching for fusion potentials appropriate for the \WZW models over the other (simply-connected) simple groups $\group{G}$, an obvious class of potentials to consider is those of the form (compare \eqnref{eqnFusPotSUSp})
\begin{equation} \label{eqnGenPot}
V_{k+\dCox}^{\Gamma} = \frac{1}{k+\dCox} \sum_{\mu \in \Gamma} e^{\brac{k+\dCox} \mu}.
\end{equation}
Here, $\Gamma$ is a finite $\group{W}$-invariant set of integral weights.  This ensures that these potentials are polynomials in the fundamental characters with rational coefficients.  Indeed, the derivatives of such polynomials have integral coefficients, as may be seen by differentiating the generating function
\begin{equation*}
\func{V^{\Gamma}}{t} = \sum_{m=1}^{\infty} \brac{-1}^{m-1} V_m^{\Gamma} t^m = \log \sqbrac{\prod_{\mu \in \Gamma} \brac{1 + e^{\mu} t}}.
\end{equation*}

In this section, we will show (with the aid of an example) that the fusion algebra of these other \WZW models is not described by potentials from this class\footnote{To be precise, we will prove that the potential cannot take the form of \eqnref{eqnGenPot} for all levels, unless $\group{G}$ is $\func{\group{SU}}{r+1}$ or $\func{\group{Sp}}{2r}$.}.  For our example, we choose the exceptional group $\group{G}_2$ because its weight space is easily visualised.  Specifically, we consider the two potentials obtained from \eqnref{eqnGenPot} by taking $\Gamma$ to be the Weyl orbit $\func{\group{W}}{\Lambda_i}$ of a fundamental weight.  One might prefer to take the potentials based on the weights of the fundamental representations, but this leads to more difficult computations.

\begin{center}
\begin{figure}
\includegraphics[width=10cm]{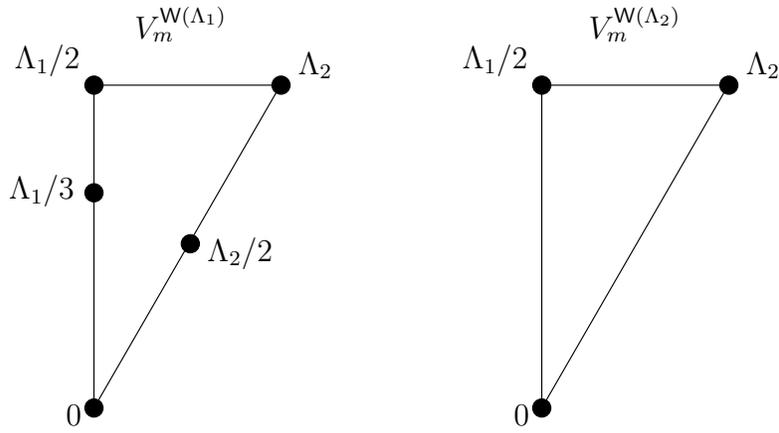}
\caption{The (shifted) critical points $\lambda + \rho$ of the potentials $V_{k+\dCox}^{\func{\group{W}}{\Lambda_1}}$ and $V_{k+\dCox}^{\func{\group{W}}{\Lambda_2}}$ for $\group{G}_2$ as a function of the weight space.  (Our convention is that $\Lambda_1$ is the highest weight of the adjoint representation.)} \label{figG2CritPts}
\end{figure}
\end{center}

As in \secref{secFusPotVerify}, we evaluate these potentials on the weight space (at $\xi_{\lambda}$).  It is extremely important to realise that as functions on the weight space, the potentials are invariant under the shifted action of the affine Weyl groups $\affine{\group{W}}_k$ \emph{for all $k$} (because the level dependence is essentially trivial).  We can therefore restrict to computing the critical points in a fundamental alcove at (effective) level $\kappa \equiv k + \dCox = 1$ (a truly fundamental domain for the periodicity of the potentials).  The results are shown in \figref{figG2CritPts}.  It is immediately evident that in contrast with the $\func{\group{SU}}{r+1}$ and $\func{\group{Sp}}{2r}$ fusion potentials, these $\group{G}_2$ potentials have critical points (with respect to the Dynkin labels $\lambda_i$) which include, but are not limited to, the weight lattice.

\psfrag{0}{$0$}
\psfrag{L1/2}{$\Lambda_1 / 2$}
\psfrag{L1/3}{$\Lambda_1 / 3$}
\psfrag{L2}{$\Lambda_2$}
\psfrag{L2/2}{$\Lambda_2 / 2$}
\psfrag{V1}[][]{$V_m^{\func{\group{W}}{\Lambda_1}}$}
\psfrag{V2}[][]{$V_m^{\func{\group{W}}{\Lambda_2}}$}
\vspace{-\baselineskip}

\begin{center}
\begin{figure}
\includegraphics[height=8cm]{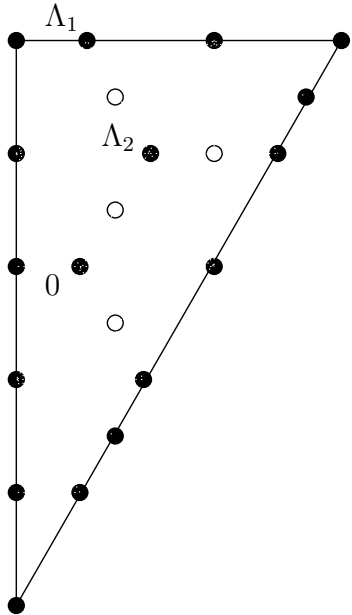}
\caption{The critical points $\lambda$ of the potential $V_{k+\dCox}^{\func{\group{W}}{\Lambda_2}}$ for $\group{G}_2$ in the shifted fundamental alcove at level $k = 1$.  The white points denote those in the interior which do not belong to the weight lattice.} \label{figG2CritPtsk=1}
\end{figure}
\end{center}

These non-integral critical points are the crux of the matter.  When these critical points lie on a shifted (level-$k$) alcove boundary, we saw in \secref{secFusPotVerify} that they need not correspond to genuine critical points (with respect to the fundamental characters).  However, any critical point in the interior of a shifted alcove is necessarily a critical point with respect to the fundamental characters, and Gepner's characterisation of the fusion variety requires these to be integral.  Unfortunately, at any given level $k > 0$, the invariance of the critical points under $\affine{\group{W}}_{k'}$ for all $k'$ means that there will always be non-integral critical points in the interior of the alcoves (for $k$ sufficiently large).  This is illustrated in \figref{figG2CritPtsk=1} for the potential $V_5^{\func{\group{W}}{\Lambda_1}}$ (corresponding to level $k = 1$).  It follows that the potentials based on the Weyl orbits of the $\group{G}_2$ fundamental weights do not describe the fusion variety.

\psfrag{0}{$0$}
\psfrag{L1}{$\Lambda_1$}
\psfrag{L2}{$\Lambda_2$}
\vspace{-\baselineskip}

We can, of course, consider potentials $V_{k+\dCox}^{\Gamma}$ based on more complicated $\group{W}$-invariant sets $\Gamma$.  However, when evaluating on the weight space, any such potential is just a $\group{W}$-invariant linear combination of formal exponentials of integral weights, and so is a polynomial in the potentials $V_{k+\dCox}^{\func{\group{W}}{\Lambda_1}}$ and $V_{k+\dCox}^{\func{\group{W}}{\Lambda_2}}$ considered before.  It follows now from the chain rule for differentiation that if $\lambda + \rho$ is a common critical point of all the $V_{k+\dCox}^{\func{\group{W}}{\Lambda_i}}$, then it is also a critical point of $V_{k+\dCox}^{\Gamma}$.  From \figref{figG2CritPts}, we see that any potential $V_{k+\dCox}^{\Gamma}$ for $\group{G}_2$ will have critical points at non-integral weights, and so will not correctly describe the fusion variety.

The situation is similarly bleak for the other simple groups because any potential of the form $V_{k+\dCox}^{\Gamma}$ will have (shifted) critical points at the vertices of the affine alcoves (at all levels).  We will demonstrate this claim shortly.  What it implies is that the only time a potential of this form stands of chance of describing the fusion variety is when the alcove vertices are integral (at all levels).  This only happens when the comarks of the Lie group are all unity, which is only the case for $\group{G} = \func{\group{SU}}{r+1}$ and $\func{\group{Sp}}{2r}$.

Let us finish with the promised demonstration.  Our earlier remarks show that it is sufficient to consider the potentials $V_m^{P_{\Lambda_i}}$, $i = 1 , \ldots , r$.  We will show that these always have critical points (with respect to $\lambda$) when $\lambda + \rho$ is the vertex of an affine alcove.  Identifying $m$ with $k+\dCox$, the condition for $V_m^{P_{\Lambda_i}}$ to have a critical point is just that $\func{J_{ij}}{-2 \pi \ii \brac{\lambda + \rho}} = 0$ for each $j$.  We therefore need to show that $\func{J}{-2 \pi \ii \nu} = 0$ whenever $\nu$ is an alcove vertex.

We rewrite \eqnref{eqnDetJw} in terms of the $i^{\text{th}}$ row of $J$, $\nabla_{\lambda} \chi_i$:
\begin{equation*}
\func{\nabla_{\lambda} \chi_i}{-2 \pi \ii \func{w}{\nu}} = \func{\nabla_{\lambda} \chi_i}{-2 \pi \ii \nu} w.
\end{equation*}
Here $w$ (on the right hand side) denotes the matrix representing $w$ with respect to the basis of fundamental weights.  We will treat the row vector $\func{\nabla_{\lambda} \chi_i}{-2 \pi \ii \nu}$ as an element of the dual of the weight space (the Cartan subalgebra).

We can also restrict our attention to the fundamental alcove vertices, by $\affine{\group{W}}$-invariance of the characters.  If $\nu = 0$, then $\nu$ is fixed by every $w \in \group{W}$, so $\func{\nabla_{\lambda} \chi_i}{-2 \pi \ii \nu}$ is a row vector fixed by every $w \in \group{W}$.  Thus, $\func{\nabla_{\lambda} \chi_i}{0}$ is the zero vector (for each $i$), verifying our claim for this vertex (and its $\affine{\group{W}}$-images).

The other fundamental alcove vertices have the form $\nu = \Lambda_j / a_j^{\vee}$, where $a_j^{\vee}$ is the $j^{\text{th}}$ comark of $\alg{g}$.  As $\nu$ is invariant under all the simple Weyl reflections except $w_j$, $\func{\nabla_{\lambda} \chi_i}{-2 \pi \ii \nu}$ is also invariant under all these simple reflections, hence $\func{\nabla_{\lambda} \chi_i}{-2 \pi \ii \nu}$ is orthogonal to every simple root except $\alpha_j$.  But, $\nu$ is fixed by the affine reflection about the hyperplane $\bilin{\mu}{\theta} = 1$.  This reflection has the form $\func{\affine{w}}{\mu} = \func{w_{\theta}}{\mu} + \theta$, where $w_{\theta} \in \group{W}$ is the Weyl reflection associated with the highest root $\theta$.  Hence, using the invariance of the characters under translations in $\group{Q}^{\vee}$,
\begin{equation*}
\func{\nabla_{\lambda} \chi_i}{-2 \pi \ii \nu} = \func{\nabla_{\lambda} \chi_i}{-2 \pi \ii \brac{\func{w_{\theta}}{\nu} + \theta}} = \func{\nabla_{\lambda} \chi_i}{-2 \pi \ii \func{w_{\theta}}{\nu}} = \func{\nabla_{\lambda} \chi_i}{-2 \pi \ii \nu} w_{\theta}.
\end{equation*}
It follows now that $\func{\nabla_{\lambda} \chi_i}{-2 \pi \ii \nu}$ is also orthogonal to $\theta$.  But, $\theta$ and the simple roots, excepting $\alpha_j$, together constitute a basis of the weight space (as the mark $a_j$ never vanishes).  Thus, $\func{\nabla_{\lambda} \chi_i}{-2 \pi \ii \nu}$ is again the zero vector, verifying our claim for all the vertices of the fundamental alcove.

\section{Presentations of Fusion Rings} \label{secFusRing}

We now turn to the study of fusion rings over $\ZZ$.  Given the results of \secref{secGenPot}, we introduce a characterisation of the fusion ideal $\fusionideal{k}{\ZZ}$ for general \WZW models which makes no mention of potentials.  We then analyse this characterisation in the cases of $\func{\group{SU}}{r+1}$ and $\func{\group{Sp}}{2r}$, and show that it can be reduced to recover the potentials of \eqnDref{eqnFusPotSU}{eqnFusPotSp}.  We would like to emphasise that this constitutes a derivation of these fusion potentials over $\ZZ$, and not an \emph{a posteriori} verification over $\CC$.  In \secref{secFusSpin}, we will apply this reduction to $\func{\group{Spin}}{2r+1}$.

\subsection{A General Characterisation} \label{secGrobner}

We begin with the simple observation that given any weight $\lambda$ and $\affine{w} \in \affine{\group{W}}_k$, we have
\begin{equation} \label{eqnObvious}
\chi_{\lambda} - \det \affine{w} \ \chi_{\affine{w} \cdot \lambda} \in \fusionideal{k}{\ZZ}.
\end{equation}
(The definition of character has been extended to non-dominant weights by Weyl's character formula.)  This follows easily from Gepner's characterisation of the fusion algebra, \propref{propGepner} (and the remarks which follow it).  Since the fusion ideal is dividing (\secref{secIntro}), it follows that $\chi_{\lambda} \in \fusionideal{k}{\ZZ}$ whenever $\lambda$ is on a shifted affine alcove boundary.

Let $L_{\lambda}$ denote the irreducible representation of $\group{G}$ of highest weight $\lambda$.  Letting $\lambda_i$ denote the Dynkin labels of the weight $\lambda$, it follows from the familiar properties of the representation ring that $\lambda$ is the highest weight of the representation $L_{\Lambda_1}^{\otimes \lambda_1} \otimes \cdots \otimes L_{\Lambda_r}^{\otimes \lambda_r}$.  As a polynomial in the character ring, $\polyring{\ZZ}{\chi_1 , \ldots , \chi_r}$, we see that the character $\chi_{\lambda}$ has the form
\begin{equation*}
\chi_{\lambda} = \chi_1^{\lambda_1} \cdots \chi_r^{\lambda_r} - \ldots
\end{equation*}
where the omitted terms correspond, in a sense, to lower weights which we regard as being of lesser importance.  Our strategy now is to make this lack of importance precise by introducing a monomial ordering on the character ring such that the leading term (\textsc{lt}) of $\chi_{\lambda}$ is precisely $\LT{\chi_{\lambda}} = \chi_1^{\lambda_1} \cdots \chi_r^{\lambda_r}$.  Of course, we are studying fusion, so we also want to assign (relative) importance to characters according to whether the associated weight is on a shifted affine alcove boundary or not.  In particular, we should distinguish weights on the boundary $\bilin{\lambda}{\theta} = k+1$ from those inside the fundamental alcove $\bilin{\lambda}{\theta} \leqslant k$.

Happily, these requirements can both be satisfied by defining a \emph{monomial ordering} $\prec$ on the character ring, $\ZZ \sqbrac{\chi_1 , \ldots , \chi_r}$, by
\begin{equation*}
\chi_1^{\lambda_1} \cdots \chi_r^{\lambda_r} \prec \chi_1^{\mu_1} \cdots \chi_r^{\mu_r} \qquad \text{if and only if}
\end{equation*}
\begin{align*}
\bilin{\lambda}{\theta} &< \bilin{\mu}{\theta}, & &\text{or} & & & & & & \\
\bilin{\lambda}{\theta} &= \bilin{\mu}{\theta} & &\text{and} & \bilin{\lambda}{\rho} &< \bilin{\mu}{\rho}, & &\text{or} & & \\
\bilin{\lambda}{\theta} &= \bilin{\mu}{\theta} & &\text{and} & \bilin{\lambda}{\rho} &= \bilin{\mu}{\rho} & &\text{and} & \chi_1^{\lambda_1} \cdots \chi_r^{\lambda_r} &\prec' \chi_1^{\mu_1} \cdots \chi_r^{\mu_r},
\end{align*}
where $\prec'$ is any other monomial ordering, lexicographic for definiteness.  This is an example of a weight order \cite{CoxIde} (and is therefore a genuine monomial ordering).

We demonstrate that $\LT{\chi_{\lambda}}$ is indeed $\chi_1^{\lambda_1} \cdots \chi_r^{\lambda_r}$.  This proceeds inductively on the height, as it is obvious when $\lambda$ is zero or a fundamental weight.  We decompose $L_{\Lambda_1}^{\otimes \lambda_1} \otimes \cdots \otimes L_{\Lambda_r}^{\otimes \lambda_r}$ into irreducible representations, so that
\begin{equation*}
\chi_1^{\lambda_1} \cdots \chi_r^{\lambda_r} = \chi_{\lambda} + \sum_{\mu} c_{\mu} \chi_{\mu},
\end{equation*}
where the $\mu$ are all of lower height than $\lambda$:  $\bilin{\mu}{\rho} < \bilin{\lambda}{\rho}$.  By induction, $\LT{\chi_{\lambda}}$ is the greatest (under $\prec$) of $\chi_1^{\lambda_1} \cdots \chi_r^{\lambda_r}$ and the monomials $- c_{\mu} \chi_1^{\mu_1} \cdots \chi_r^{\mu_r}$.  Now, since each $\mu$ is a weight of $L_{\Lambda_1}^{\otimes \lambda_1} \otimes \cdots \otimes L_{\Lambda_r}^{\otimes \lambda_r}$, $\mu = \lambda - \sum_i m_i \alpha_i$, where the $m_i$ are non-negative integers.  It follows that $\bilin{\mu}{\theta} \leqslant \bilin{\lambda}{\theta}$ since the Dynkin labels of $\theta$ are never negative.  But, in the definition of $\prec$, ties in $\bilin{\cdot}{\theta}$ are broken by height, hence $\chi_1^{\lambda_1} \cdots \chi_r^{\lambda_r}$ is the greatest of the monomials (under $\prec$) as required.

Consider now the ideal $\ideal{\LT{\fusionideal{k}{\ZZ}}}$ generated by the leading terms (with respect to $\prec$) of the polynomials in the fusion ideal.  Since the fusion ring is freely generated (as a $\ZZ$-module) by (the cosets of) the characters of the weights in $\affine{\group{P}}_k$, the leading terms $\chi_1^{\lambda_1} \cdots \chi_r^{\lambda_r}$, with $\bilin{\lambda}{\theta} \leqslant k$ must be the only monomials not in $\ideal{\LT{\fusionideal{k}{\ZZ}}}$.  That is, $\ideal{\LT{\fusionideal{k}{\ZZ}}}$ is freely generated as an abelian group by the set of monomials $\mathcal{M} = \set{\chi_1^{\lambda_1} \cdots \chi_r^{\lambda_r} \colon \bilin{\lambda}{\theta} > k}$.  

As an ideal, it is now easy to see that $\ideal{\LT{\fusionideal{k}{\ZZ}}}$ is generated by the \emph{atomic} monomials of $\mathcal{M}$, where the atomic monomials are defined to be those which \emph{cannot} be expressed as the product of a fundamental character and a monomial from $\mathcal{M}$.  Equivalently, atomic monomials are those corresponding to weights from which one cannot subtract any fundamental weight and still remain in the set of weights corresponding to $\mathcal{M}$.

\begin{center}
\begin{figure}
\includegraphics[width=13cm]{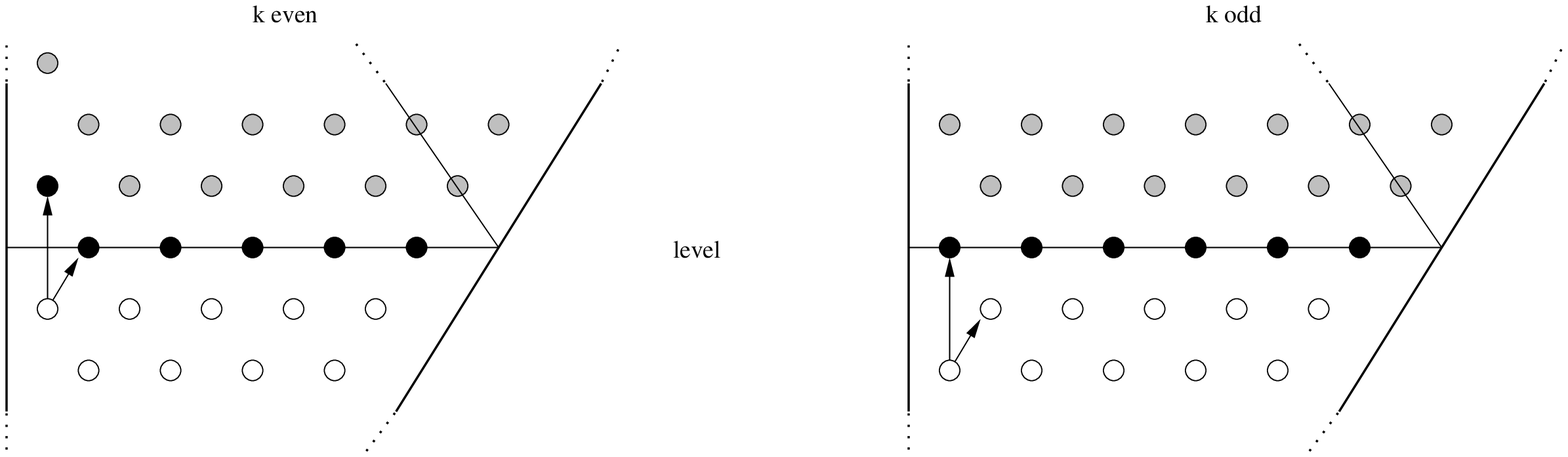}
\caption{The weights corresponding to the atomic monomials for the ideal $\ideal{\LT{\fusionideal{k}{\ZZ}}}$ associated with $\group{G}_2$ at even and odd level.  Weights corresponding to monomials in the ideal are grey or black, the latter corresponding to atomic monomials.  The arrows indicate the effect of multiplying by $\chi_1$ and $\chi_2$.} \label{figG2AtomMon}
\end{figure}
\end{center}

It should be clear that every weight $\lambda$ with $\bilin{\lambda}{\theta} = k+1$ corresponds to an atomic monomial.  In fact, for $\func{\group{SU}}{r+1}$ and $\func{\group{Sp}}{2r}$, these are all the atomic monomials, as the comarks are $a_i^{\vee} = 1$ (so if $\bilin{\mu}{\theta} > k+1$, one can always subtract a fundamental weight from $\mu$ yet remain in $\mathcal{M}$).  For other groups, it will generally be necessary to include other monomials.  For example, $a_1^{\vee} = 2$ for $\group{G}_2$, so it follows that when the level $k$ is even, the monomial $\chi_1^{\brac{k+2} / 2}$ is also atomic (this is illustrated in \figref{figG2AtomMon}).

\psfrag{level}[][]{$\bilin{\lambda}{\theta} = k+1$}
\psfrag{k even}[][]{$k$ even}
\psfrag{k odd}[][]{$k$ odd}
\vspace{-\baselineskip}

Let $\chi_1^{\lambda_1} \cdots \chi_r^{\lambda_r}$ be an atomic monomial of $\mathcal{M}$.  If the associated weight $\lambda$ is on a shifted affine alcove boundary, we associate to this atomic monomial the polynomial $p_{\lambda} = \chi_{\lambda} \in \fusionideal{k}{\ZZ}$.  If not, we use \eqnref{eqnObvious} to reflect $\lambda$ into the fundamental affine alcove, and take $p_{\lambda} = \chi_{\lambda} - \det \affine{w} \ \chi_{\affine{w} \cdot \lambda} \in \fusionideal{k}{\ZZ}$.  In either case, we have constructed a $p_{\lambda}$ in the fusion ideal whose leading term with respect to $\prec$ is $\chi_1^{\lambda_1} \cdots \chi_r^{\lambda_r}$.  Therefore,
\begin{align*}
\ideal{\LT{\fusionideal{k}{\ZZ}}} &= \ideal{\text{atomic $\chi_1^{\lambda_1} \cdots \chi_r^{\lambda_r}$ in $\mathcal{M}$}} \\
&= \ideal{\LT{p_{\lambda}} \colon \text{$\lambda$ is associated to an atomic monomial in $\mathcal{M}$}}.
\end{align*}
But, this is exactly the definition of a \emph{Gr\"{o}bner basis} for $\fusionideal{k}{\ZZ}$ \cite{CoxIde,CoxUsi}.

\begin{proposition} \label{propGrobner}
The polynomials $p_{\lambda}$ constructed above for each weight $\lambda$ associated to an atomic monomial of $\mathcal{M} = \set{\chi_1^{\lambda_1} \cdots \chi_r^{\lambda_r} \colon \bilin{\lambda}{\theta} > k}$ form a Gr\"{o}bner basis for the fusion ideal $\fusionideal{k}{\ZZ}$, with respect to the monomial ordering $\prec$.  That is,
\begin{equation*}
\fusionideal{k}{\ZZ} = \ideal{p_{\lambda} \colon \text{$\lambda$ is associated to an atomic monomial in $\mathcal{M}$}}.
\end{equation*}
\end{proposition}

Note the crucial, but subtle, r\^{o}le played by the monomial ordering $\prec$.  Note also that because the Gr\"{o}bner basis given has elements whose leading coefficient is unity, this presentation shows explicitly that the fusion ideal is dividing.  Whilst this presentation has a nice Lie-theoretic interpretation, it is rather more cumbersome than we would wish for.  Indeed, a presentation in terms of a potential would give a set of $r = \rank \group{G}$ generators for the fusion ideal (at every level $k$), whereas \propref{propGrobner} gives a set whose cardinality is of the order of $k^{r-1}$.  We will therefore indicate in what follows how one can reduce the number of generators to something a bit more manageable (at least for the classical groups).

\subsection{Deriving Fusion Potentials} \label{secDerFusPot}

We will begin with the case of $\func{\group{SU}}{r+1}$.  As noted in \secref{secGrobner}, the atomic monomials of $\mathcal{M} = \set{\chi_1^{\lambda_1} \cdots \chi_r^{\lambda_r} \colon \bilin{\lambda}{\theta} > k}$ are precisely those corresponding to weights $\lambda$ with $\bilin{\lambda}{\theta} = k+1$.  It follows from \propref{propGrobner} that
\begin{equation*}
\fusionideal{k}{\ZZ} = \ideal{\chi_{\lambda} \colon \bilin{\lambda}{\theta} = k+1}.
\end{equation*}
The highest root has the form $\theta = \eps_1 - \eps_{r+1}$, so for these weights, $k+1 = \bilin{\lambda}{\theta} = \lambda^1 - \lambda^{r+1} = \lambda^1$.  Here, we write $\lambda = \sum_{j=1}^{r+1} \lambda^j \eps_j$, and fix the ambiguity corresponding to $\sum_{j=1}^{r+1} \eps_j = 0$ by setting $\lambda^{r+1} = 0$.  We emphasise that the $\lambda^j$ are not to be confused with the Dynkin labels $\lambda_j$.

We now use the \emph{Jacobi-Trudy identity}, \eqnref{eqnJTA}, to decompose these generators of the fusion ideal into complete symmetric polynomials (denoted by $H_m$) in the $q_i$.  We have
\begin{equation*}
\chi_{\lambda} = 
\begin{vmatrix}
H_{\lambda^1} & H_{\lambda^2 - 1} & \cdots & H_{\lambda^r - r + 1} \\
H_{\lambda^1 + 1} & H_{\lambda^2} & \cdots & H_{\lambda^r - r + 2} \\
\vdots & \vdots & \ddots & \vdots \\
H_{\lambda^1 + r - 1} & H_{\lambda^2 + r - 2} & \cdots & H_{\lambda^r}
\end{vmatrix}
 = 
\begin{vmatrix}
H_{k+1} & H_{\lambda^2 - 1} & \cdots & H_{\lambda^r - r + 1} \\
H_{k+2} & H_{\lambda^2} & \cdots & H_{\lambda^r - r + 2} \\
\vdots & \vdots & \ddots & \vdots \\
H_{k+r} & H_{\lambda^2 + r - 2} & \cdots & H_{\lambda^r}
\end{vmatrix}
.
\end{equation*}
Since $H_m = \chi_{m \Lambda_1} \in \polyring{\ZZ}{\chi_1 , \ldots , \chi_r}$, expanding this determinant down the first column gives $\chi_{\lambda}$ as a $\polyring{\ZZ}{\chi_1 , \ldots , \chi_r}$-linear combination of the $H_{k+i} = \chi_{\brac{k+i} \Lambda_1}$, where $i = 1 , \ldots , r$.  Therefore,
\begin{equation*}
\fusionideal{k}{\ZZ} \subseteq \ideal{\chi_{\brac{k+i} \Lambda_1} \colon i = 1 , \ldots , r}.
\end{equation*}
Conversely, we show that each $\brac{k+i} \Lambda_1$, $i = 1 , \ldots , r$, is on a shifted affine alcove boundary, hence is fixed by an affine reflection $\affine{w}$, and thus that $\chi_{\brac{k+i} \Lambda_1}$ is in the fusion ideal.  This amounts to verifying that $\bilin{\brac{k+i} \Lambda_1}{\alpha} \in \brac{k + \dCox} \ZZ$ for some root $\alpha$, and the reader can easily check that $\alpha = \eps_1 - \eps_{r+2-i}$ works.  We have therefore demonstrated that
\begin{equation} \label{eqnFusGenSU}
\fusionideal{k}{\ZZ} = \ideal{\chi_{\brac{k+i} \Lambda_1} \colon i = 1 , \ldots , r}.
\end{equation}

It is rather pleasing that such a simple device can reduce the number of generators from (the order of) $k^{r-1}$ to $r$.  Before turning to the integration of these generators to a potential, we would like to mention one further observation that may be of interest.  We consider the characters $\chi_{k \Lambda_1 + \Lambda_i}$, where $i = 1 , \ldots , r$.  Expanding with the Jacobi-Trudy identity, we find that
\begin{align*}
\chi_{k \Lambda_1 + \Lambda_1} &= H_{k+1} \\
\chi_{k \Lambda_1 + \Lambda_2} &= H_1 H_{k+1} - H_{k+2} \\
\chi_{k \Lambda_1 + \Lambda_3} &= \brac{H_1^2 - H_2} H_{k+1} - H_1 H_{k+2} + H_{k+3} \\
\chi_{k \Lambda_1 + \Lambda_4} &= \brac{H_1^3 - 2 H_1 H_2 + H_3} H_{k+1} - \brac{H_1^2 - H_2} H_{k+2} + H_1 H_{k+3} - H_{k+4} \\
&\mspace{9mu} \vdots
\end{align*}
We call this the \emph{method of $1$'s} due to the line of $1$'s which appear off-diagonal in the Jacobi-Trudy expansion of these characters.  These equations show (inductively) that there is another simple generating set for the fusion ideal:
\begin{equation*}
\fusionideal{k}{\ZZ} = \ideal{\chi_{k \Lambda_1 + \Lambda_i} \colon i = 1 , \ldots , r}.
\end{equation*}
This generating set is suggested by the computations of \cite{FreBra} (though not explicitly stated there) on the corresponding brane charge groups\footnote{To elaborate somewhat, the authors of \cite{FreBra} computed the brane charge group of the level $k$ $\func{\group{SU}}{r+1}$ \WZW model from the greatest common divisor of the dimensions of the irreducible representations of highest weight $k \Lambda_1 + \Lambda_i$, $i = 1 , \ldots , r$.  In \cite{BouDBr02}, the brane charge group was shown to be determined by the greatest common divisor of the dimensions of any set of generators of the ideal $\fusionideal{k}{\ZZ}$ of the fusion ring.  This suggests that the $\chi_{k \Lambda_1 + \Lambda_i}$ are such a set of generators, and here we have given a simple proof of this fact.}.  Note that this set has the nice property of consisting entirely of characters $\chi_{\lambda}$ with $\bilin{\lambda}{\theta} = k+1$.

We now turn to the derivation of the fusion potential, \eqnref{eqnFusPotSU}.  Let $E_n$ denote the $n^{\text{th}}$ elementary symmetric polynomial in the $q_i$.  From the identity $\sum_m H_m t^m = \big[ \sum_n \brac{-1}^n E_n t^n \big] ^{-1}$, we can derive
\begin{equation} \label{eqnDHDE}
\parD{H_m}{E_j} = \brac{-1}^{j+1} \sum_n H_n H_{m-j-n}.
\end{equation}
For $\func{\group{SU}}{r+1}$, $E_j = \chi_j \equiv \chi_{\Lambda_j}$ for $j = 1 , \ldots , r$, so we see that
\begin{equation*}
\brac{-1}^{i-1} \parD{H_{k+\dCox-i}}{\chi_j} = \brac{-1}^{i+j} \sum_n H_n H_{k+\dCox-i-j-n}
\end{equation*}
is symmetric in $i$ and $j$.  Therefore, $\sum_i \brac{-1}^{i-1} H_{k+\dCox-i} \dd \chi_i$ is a closed $1$-form, hence integrates to a potential $V_{k+\dCox}$ (there is no topology).

We can compute this potential using generating functions.  If $\func{V}{t} = \sum_m \brac{-1}^{m-1} V_m t^m$, then
\begin{align*}
\parD{\func{V}{t}}{\chi_i} &= \sum_m \brac{-1}^{m+i} H_{m-i} t^m = \frac{t^i}{\prod_{\ell} \brac{1 + q_{\ell} t}} = \frac{t^i}{\sum_n E_n t^n} \\
&= \frac{t^i}{1 + \chi_1 t + \ldots + \chi_r t^r + t^{r+1}} \\
\Rightarrow \qquad \func{V}{t} &= \log \sqbrac{1 + \chi_1 t + \ldots + \chi_r t^r + t^{r+1}},
\end{align*}
up to a constant.  This is of course \eqnref{eqnFusPotGFSU}, from which one can easily recover the fusion potential, \eqnref{eqnFusPotSU}.

We would like to emphasise once again that not only have we given a complete derivation of the fusion potential for the $\func{\group{SU}}{r+1}$ \WZW models, but we have shown that this potential describes the fusion process over $\ZZ$, rather than just over $\CC$.

Consider now the fusion ring for $\func{\group{Sp}}{2r}$.  As before, \propref{propGrobner} gives the characters $\chi_{\lambda}$ with $\bilin{\lambda}{\theta} = k+1$ as a set of generators for the fusion ideal, $\fusionideal{k}{\ZZ}$.  The highest root is $\theta = 2 \eps_1$, so for these characters, $k+1 = \bilin{\lambda}{\theta} = \lambda^1$ (note that $\norm{\eps_i}^2 = \frac{1}{2}$).  We expand the $\func{\group{Sp}}{2r}$ Jacobi-Trudy identity, \eqnref{eqnJTC}, down the first column.  Noting that $H_m = \chi_{m \Lambda_1}$, this shows that the generating characters can be expressed as $\polyring{\ZZ}{\chi_1 , \ldots , \chi_r}$-linear combinations of the $r$ elements $H_{k+1}$ and $H_{k+1+i} + H_{k+1-i}$ ($i = 1 , \ldots , r-1$).  Here, the $H_m$ are complete symmetric polynomials in the $q_i$ and their inverses.  It is obvious that these elements belong to $\fusionideal{k}{\ZZ}$, hence
\begin{equation} \label{eqnFusGenSp}
\fusionideal{k}{\ZZ} = \ideal{\chi_{\brac{k+1} \Lambda_1} , \chi_{\brac{k+1+i} \Lambda_1} + \chi_{\brac{k+1-i} \Lambda_1} \colon i = 1 , \ldots , r-1}.
\end{equation}
Applying the method of $1$'s to these elements gives an alternative set of generators:
\begin{equation*}
\fusionideal{k}{\ZZ} = \ideal{\chi_{k \Lambda_1 + \Lambda_i} \colon i = 1 , \ldots , r}.
\end{equation*}

Deriving a potential from these generators is somewhat more cumbersome than before.  For this purpose, we use the set of generators
\begin{equation*}
\set{\sum_{\ell = 0}^{r-i} H_{k+\dCox-i-2 \ell} \colon i = 1 , \ldots , r},
\end{equation*}
which is easily derived from those given above.  From \eqnref{eqnDHDE} and the expressions for $E_n$ in terms of the $\chi_j$ \cite{FulRep}, we compute that
\begin{equation*}
\brac{-1}^{i-1} \parD{}{\chi_j} \sum_{\ell = 0}^{r-i} H_{k+\dCox-i-2 \ell} = \brac{-1}^{i+j} \sum_n H_n \sum_{m=0}^{r-i} \sum_{m'=0}^{r-j} H_{k+\dCox-n-i-j-2 \brac{m+m'}},
\end{equation*}
which is symmetric in $i$ and $j$ (indeed, this symmetry is what suggests the above generating set, as it leads to a closed $1$-form).  These generators may therefore be integrated to a potential, and the derivation may be completed using generating functions as in the $\func{\group{SU}}{r+1}$ case.  In this way, we recover \eqnref{eqnFusPotGFSp} and therefore the fusion potential, \eqnref{eqnFusPotSp}.

\section{Presentations for $\func{\group{Spin}}{2r+1}$} \label{secFusSpin}

We now apply the techniques of \secref{secGrobner} to the fusion rings of the \WZW models over $\func{\group{Spin}}{2r+1}$.  We are not aware of any concise, representation-theoretic presentations of these rings (nor of the corresponding algebras) in the literature\footnote{In the course of preparing this section, we were made aware of a conjecture regarding the presentations of the fusion ideals of the $\func{\group{Spin}}{2r+1}$ (and $\func{\group{Spin}}{2r}$) \WZW models \cite{Boysal}.  This elegant conjecture amounts to the statement that the fusion ideal at level $k$ is the radical of the ideal generated by the $\chi_{\brac{k+i} \Lambda_1}$, for $i = 1 , 2 , \ldots , \dCox - 1$.  This is a generalisation of the $\func{\group{SU}}{r+1}$ result, \eqnref{eqnFusGenSU}.  It is further conjectured that the radical of this ideal is generated by the above characters and $\chi_{k \Lambda_1 + \Lambda_r}$ ($\chi_{k \Lambda_1 + \Lambda_{r-1}}$ is also needed for $\func{\group{Spin}}{2r}$).}.  We will see that the appropriate Jacobi-Trudy identities may be employed to substantially simplify the presentations given by \propref{propGrobner}, though the simplification turns out to be not quite so drastic as that found for $\func{\group{SU}}{r+1}$ and $\func{\group{Sp}}{2r}$.  In particular, it seems rather doubtful that the presentations obtained are related to potentials.

Recall from \secref{secGrobner} that we can derive a generating set for the fusion ideal $\fusionideal{k}{\ZZ}$ by computing the atomic monomials of the set $\set{\chi_1^{\lambda_1} \cdots \chi_r^{\lambda_r} \colon \bilin{\lambda}{\theta} > k}$.  As shown there for $\group{G}_2$, this computation depends upon the comarks $a_i^{\vee}$, which for $\func{\group{Spin}}{2r+1}$ are $1$ for $i = 1 , r$, and $2$ otherwise (we will only consider $r > 2$).  The atomic monomials therefore correspond to the weights
\begin{align*}
\text{$k$ odd} &: & &\set{\lambda \colon \bilin{\lambda}{\theta} = k+1} \\
\text{$k$ even} &: & &\set{\lambda \colon \bilin{\lambda}{\theta} = k+1} \cup \set{\lambda \colon \bilin{\lambda}{\theta} = k+2 \text{ and } \lambda_1 = \lambda_r = 0}.
\end{align*}
Finding elements of $\fusionideal{k}{\ZZ}$ whose leading terms are these monomials is easy, and we deduce from \propref{propGrobner} that the fusion ring is generated by:
\begin{align} \label{eqnGrobGenB}
\begin{split}
\text{$k$ odd}: \qquad &\set{\chi_{\lambda} \colon \bilin{\lambda}{\theta} = k+1} \\
\text{$k$ even}: \qquad &\set{\chi_{\lambda} \colon \bilin{\lambda}{\theta} = k+1} \cup \set{\chi_{\lambda} + \chi_{\lambda - \theta} \colon \bilin{\lambda}{\theta} = k+2 \text{ and } \lambda_1 = \lambda_r = 0}.
\end{split}
\end{align}
We note that if $\lambda_2 = 0$, $\chi_{\lambda - \theta} = 0$.

In order to reduce the size of this generating set, we again turn to the appropriate Jacobi-Trudy identities.  As noted in \appref{secJTB}, these identities distinguish between \emph{tensor} and \emph{spinor} representations (whose highest weight $\lambda$ has $\lambda_r$ even and odd, respectively).  We consider first the tensor representations.  The appropriate Jacobi-Trudy identity, \eqnref{eqnJTBT}, gives the irreducible characters as a determinant of an $r \times r$ matrix:
\begin{equation} \label{eqnJTBT+}
\chi_{\lambda} = 
\begin{vmatrix}
H_{\lambda^1} - H_{\lambda^1 - 2} & H_{\lambda^2 - 1} - H_{\lambda^2 - 3} & \cdots & H_{\lambda^r + 1 - r} - H_{\lambda^r - 1 - r} \\
H_{\lambda^1 + 1} - H_{\lambda^1 - 3} & H_{\lambda^2} - H_{\lambda^2 - 4} & \cdots & H_{\lambda^r + 2 - r} - H_{\lambda^r - 2 - r} \\
\vdots & \vdots & \ddots & \vdots \\
H_{\lambda^1 + r - 1} - H_{\lambda^1 - r - 1} & H_{\lambda^2 + r - 2} - H_{\lambda^2 - r - 2} & \cdots & H_{\lambda^r} - H_{\lambda^r - 2r}
\end{vmatrix}
.
\end{equation}
Here, $\lambda^j$ denotes the components of $\lambda$ with respect to the usual orthonormal basis $\eps_j$ of the weight space, and $H_m$ denotes the $m^{\text{th}}$ complete symmetric polynomial in the $q_i = \func{\exp}{\eps_i}$, their inverses, and $1$.

How this treatment differs from the analysis of \secref{secDerFusPot}, and is thereby significantly complicated, is that $\theta = \eps_1 + \eps_2$, so $\bilin{\lambda}{\theta} = \lambda_1 + \lambda_2$.  It follows that the elements in any single column of the Jacobi-Trudy determinant of a character $\chi_{\lambda}$ with $\bilin{\lambda}{\theta} = k+1$ will not generally belong to the fusion ideal, so expanding the determinant down a single column is pointless.  Instead, we notice that the top-left $2 \times 2$ subdeterminant is the character $\chi_{\lambda^1 \eps_1 + \lambda^2 \eps_2}$, and that $\bilin{\lambda}{\theta} = k+1$ implies that \emph{this} subdeterminant is in $\fusionideal{k}{\ZZ}$.

This observation suggests that we must expand \eqnref{eqnJTBT+} down the first two columns.  In this way, $\chi_{\lambda}$ is expressed as a $\polyring{\ZZ}{\chi_1 , \ldots , \chi_r}$-linear combination of the $2 \times 2$ determinants
\begin{equation*}
\func{\psi_{m_1 m_2}}{\lambda^1 , \lambda^2} = 
\begin{vmatrix}
H_{\lambda^1 + m_1 - 1} - H_{\lambda^1 - m_1 - 1} & H_{\lambda^2 + m_1 - 2} - H_{\lambda^2 - m_1 - 2} \\
H_{\lambda^1 + m_2 - 1} - H_{\lambda^1 - m_2 - 1} & H_{\lambda^2 + m_2 - 2} - H_{\lambda^2 - m_2 - 2}
\end{vmatrix}
.
\end{equation*}
Here, $1 \leqslant m_1 < m_2 \leqslant r$ counts the $\binom{r}{2}$ choices of rows used in these subdeterminants.  We have already noted that $\func{\psi_{12}}{\lambda^1 , \lambda^2} \in \fusionideal{k}{\ZZ}$ when $\lambda^1 + \lambda^2 = k+1$, so it is natural to enquire if the same is true for general $m_1$ and $m_2$.

To investigate this, we need to digress a little in order to derive a more amenable form for the $\func{\psi_{12}}{\lambda^1 , \lambda^2}$ (\eqnref{eqnGenBT} below).  This derivation is an exercise in manipulating generating functions.  Introducing parameters $t_1$ and $t_2$, we compute
\begin{equation} \label{eqnGFBT}
\sum_{\lambda^1 , \lambda^2 \in \ZZ} \func{\psi_{m_1 m_2}}{\lambda^1 , \lambda^2} t_1^{\lambda^1} t_2^{\lambda^2} = \sum_{\lambda^1 \in \ZZ} H_{\lambda^1} t_1^{\lambda^1} \sum_{\lambda^2 \in \ZZ} H_{\lambda^2} t_2^{\lambda^2} \Det{t_j^{j-m_i} - t_j^{j+m_i}}_{i,j = 1}^2.
\end{equation}
Denoting the determinant on the right by $A_{m_1 m_2}$, we form the generating function
\begin{equation*}
\sum_{m_1 , m_2 = 0}^{\infty} A_{m_1 m_2} z_1^{m_1} z_2 ^{m_2} = \Det{\frac{\brac{t_j^{j-1} - t_j^{j+1}} z_i}{\brac{1 - t_j z_i} \brac{1 - t_j^{-1} z_i}}}_{i,j = 1}^2.
\end{equation*}
Applying \eqnref{eqnCauchy2} to this determinant gives
\begin{align*}
\sum_{m_1 , m_2 = 0}^{\infty} A_{m_1 m_2} z_1^{m_1} z_2 ^{m_2} &= \brac{1 - t_1^2} \brac{t_2 - t_2^3} \frac{\Det{\brac{t_j + t_j^{-1}}^{2-i}} \Det{\brac{z_i + z_i^{-1}}^{j-1} z_i^2}}{\displaystyle \prod_{i,j = 1}^2 \brac{1 - t_j z_i} \brac{1 - t_j^{-1} z_i}} \\
&= -A_{12} 
\begin{vmatrix}
z_1^2 & z_1^3 + z_1 \\
z_2^2 & z_2^3 + z_2
\end{vmatrix}
\prod_i \sqbrac{\sum_{m_i \in \ZZ} \func{h_{m_i}}{t_1 , t_1^{-1} , t_2 , t_2^{-1}} z_i^{m_i}},
\end{align*}
where we recognise $A_{12} = \brac{1 - t_1^2} \brac{1 - t_2^2} \brac{1 - t_1 t_2} \brac{1 - t_1^{-1} t_2}$.  Here, $h_m$ denotes the $m^{\text{th}}$ complete symmetric polynomial in the $t_i$ and their inverses (to be distinguished from the $H_m$).

It follows that $A_{12}$ is a factor of $A_{m_1 m_2}$:
\begin{equation*}
A_{m_1 m_2} = A_{12} 
\begin{vmatrix}
h_{m_2 - 2} & h_{m_2 - 1} + h_{m_2 - 3} \\
h_{m_1 - 2} & h_{m_1 - 1} + h_{m_1 - 3}
\end{vmatrix}
.
\end{equation*}
Fascinatingly, if we set $t_j = \func{\exp}{\eta_j}$, where $\eta_j$ denotes the usual orthogonal basis vectors for the weight space of $\func{\group{Sp}}{4}$, then comparing with \eqnref{eqnJTC} gives
\begin{equation*}
\frac{A_{m_1 m_2}}{A_{12}} = \chi_{\brac{m_2 - 2} \eta_1 + \brac{m_1 - 1} \eta_2}^{\func{\group{Sp}}{4}}.
\end{equation*}
This rather unexpected relation turns out to be extremely useful.  For example, we can substitute it back into \eqnref{eqnGFBT} to recover an expression for the original determinants:
\begin{equation} \label{eqnGenBT}
\func{\psi_{m_1 m_2}}{\lambda^1 , \lambda^2} = \sum_{\mu} \chi_{\brac{\lambda^1 - \mu^1} \eps_1 + \brac{\lambda^2 - \mu^2} \eps_2}^{\func{\group{Spin}}{2r+1}}.
\end{equation}
Here, the sum is over the weights $\mu = \mu^1 \eta_1 + \mu^2 \eta_2$ of the irreducible $\func{\group{Sp}}{4}$-module of highest weight $\brac{m_2 - 2} \eta_1 + \brac{m_1 - 1} \eta_2$.

Recall that the fusion ideal is generated by the characters $\chi_{\lambda}$ with $\bilin{\lambda}{\theta} = k+1$ and, if $k$ is even, by the same set augmented by the $\chi_{\lambda} + \chi_{\lambda - \theta}$ with $\bilin{\lambda}{\theta} = k+2$ and $\lambda_1 = \lambda_r = 0$.  We have seen that when the characters correspond to tensor representations, the generators of the first type may be expressed as a $\polyring{\ZZ}{\chi_1 , \ldots , \chi_r}$-linear combination of the $\func{\psi_{m_1 m_2}}{\lambda^1 , \lambda^2}$, with $\lambda^1 + \lambda^2 = k+1$.  Since $\theta = \eps_1 + \eps_2$, it follows that the Jacobi-Trudy determinant for $\chi_{\lambda}$ and $\chi_{\lambda - \theta}$ will be identical in columns $3 , \ldots , r$.  Therefore, the generators of the second type (which always correspond to tensor representations) may be expressed as a $\polyring{\ZZ}{\chi_1 , \ldots , \chi_r}$-linear combination of the $\func{\psi_{m_1 m_2}}{\lambda^1 , \lambda^2} + \func{\psi_{m_1 m_2}}{\lambda^1 - 1, \lambda^2 - 1}$, with $\lambda^1 + \lambda^2 = k+2$.  Indeed, $\lambda_1 = 0$ implies that $\lambda^1 = \lambda^2$, so the generators of the second type can \emph{all} be expressed in terms of the elements $\func{\psi_{m_1 m_2}}{\frac{k}{2} + 1 , \frac{k}{2} + 1 } + \func{\psi_{m_1 m_2}}{\frac{k}{2} , \frac{k}{2}}$.

Consider now a single $\func{\group{Spin}}{2r+1}$-character in the sum of \eqnref{eqnGenBT}, labelled by the weight $\brac{\lambda^1 - \mu^1} \eps_1 + \brac{\lambda^2 - \mu^2} \eps_2$, with $\lambda^1 + \lambda^2 = k+1$.  We can pair it with the character labelled by the weight $\brac{\lambda^1 + \mu^2} \eps_1 + \brac{\lambda^2 + \mu^1} \eps_2$, its image under the fundamental affine Weyl reflection $\affine{w}_0$.  If this character is also (always) in the sum, then we can conclude that the right-hand-side of \eqnref{eqnGenBT} belongs to $\fusionideal{k}{\ZZ}$, that is $\func{\psi_{m_1 m_2}}{\lambda^1 , \lambda^2} \in \fusionideal{k}{\ZZ}$.

But this follows immediately from the fact that the transformation
\begin{equation*}
- \mu^1 \eta_1 - \mu^2 \eta_2 \longmapsto \mu^2 \eta_1 + \mu^1 \eta_2
\end{equation*}
is precisely the action of the $\func{\group{Sp}}{4}$-Weyl reflection about the (short) root $\eta_1 + \eta_2$.  Since the sum in \eqnref{eqnGenBT} is over the weights of an $\func{\group{Sp}}{4}$-representation, which is invariant under this (indeed any) $\func{\group{Sp}}{4}$-Weyl reflection, it is clear that $\func{\psi_{m_1 m_2}}{\lambda^1 , \lambda^2} \in \fusionideal{k}{\ZZ}$ (when $\lambda^1 + \lambda^2 = k+1$).  More generally, an almost identical argument shows that $\func{\psi_{m_1 m_2}}{\frac{k}{2} + 1 , \frac{k}{2} + 1 } + \func{\psi_{m_1 m_2}}{\frac{k}{2} , \frac{k}{2}} \in \fusionideal{k}{\ZZ}$.  It follows that the generators of $\fusionideal{k}{\ZZ}$ that correspond to tensor representations can be replaced by
\begin{align*}
&\func{\psi_{m_1 m_2}}{\lambda^1 , \lambda^2}, & &\lambda^1 + \lambda^2 = k+1, \\
\text{and } &\func{\psi_{m_1 m_2}}{\frac{k}{2} + 1 , \frac{k}{2} + 1 } + \func{\psi_{m_1 m_2}}{\frac{k}{2} , \frac{k}{2}} & &\text{if $k$ is even,}
\end{align*}
where $1 \leqslant m_1 < m_2 \leqslant r$.

The story for the spinor representations ($\lambda^j$ half-integral) is much the same.  Using the appropriate Jacobi-Trudy identity, \eqnref{eqnJTBS}, we find that the $\chi_{\lambda}$ are $\polyring{\ZZ}{\chi_1 , \ldots , \chi_r}$-linear combinations of the subdeterminants
\begin{equation*}
\func{\varphi_{m_1 m_2}}{\lambda^1 , \lambda^2} = \chi_r 
\begin{vmatrix}
H_{\lambda^1 + m_1 - \frac{3}{2}} - H_{\lambda^1 - m_1 - \frac{1}{2}} & H_{\lambda^2 + m_1 - \frac{5}{2}} - H_{\lambda^2 - m_1 - \frac{3}{2}} \\
H_{\lambda^1 + m_2 - \frac{3}{2}} - H_{\lambda^1 - m_2 - \frac{1}{2}} & H_{\lambda^2 + m_2 - \frac{5}{2}} - H_{\lambda^2 - m_2 - \frac{3}{2}}
\end{vmatrix}
.
\end{equation*}
Constructing generating functions as before, one can prove that
\begin{equation} \label{eqnGenBS}
\func{\varphi_{m_1 m_2}}{\lambda^1 , \lambda^2} = \sum_{\nu} \chi_{\brac{\lambda^1 - \nu^1 - \frac{1}{2}} \eps_1 + \brac{\lambda^2 - \nu^2 - \frac{1}{2}} \eps_2 + \Lambda_r}^{\func{\group{Spin}}{2r+1}},
\end{equation}
where this sum is over the weights $\nu = \nu^1 \zeta_1 + \nu^2 \zeta_2$ of the irreducible $\func{\group{Spin}}{5}$-module of highest weight $\brac{m_2 - 2} \zeta_1 + \brac{m_1 - 1} \zeta_2$ (and the $\zeta_i$ are the usual orthonormal basis vectors for this weight space).  As before, it now follows quickly from the fact that $\zeta_1 + \zeta_2$ is a root of $\func{\group{Spin}}{5}$ that $\func{\varphi_{m_1 m_2}}{\lambda^1 , \lambda^2} \in \fusionideal{k}{\ZZ}$.

These manipulations for the tensor and spinor representations finally prove that the fusion ideal has the following generators:
\begin{align}
\text{$k$ odd} &: & \fusionideal{k}{\ZZ} &= \Bigl\langle \func{\psi_{m_1 m_2}}{\lambda^1 , \lambda^2} , \func{\varphi_{m_1 m_2}}{\lambda^1 , \lambda^2} \colon \lambda^1 + \lambda^2 = k+1, \ 1 \leqslant m_1 < m_2 \leqslant r \Bigr\rangle, \notag \\
\text{$k$ even} &: & \fusionideal{k}{\ZZ} &= \Bigl\langle \func{\psi_{m_1 m_2}}{\lambda^1 , \lambda^2} , \func{\psi_{m_1 m_2}}{\textstyle \frac{k}{2} + 1 , \frac{k}{2} + 1 } + \func{\psi_{m_1 m_2}}{\textstyle \frac{k}{2} , \frac{k}{2}} , \func{\varphi_{m_1 m_2}}{\lambda^1 , \lambda^2} \Bigr. \label{eqnFusRingGenB} \\
& & & \mspace{300mu} \Bigl. \colon \lambda^1 + \lambda^2 = k+1, \ 1 \leqslant m_1 < m_2 \leqslant r \Bigr\rangle .\notag
\end{align}
Since $\lambda^1 \geqslant \lambda^2$ are integers and half-integers in the $\psi_{m_1 m_2}$ and $\varphi_{m_1 m_2}$ respectively, it follows that the number of generators in this set is of the order of $k \binom{r}{2}$.  This compares favourably with the set of generators given in \eqnref{eqnGrobGenB}, whose number is of the order $k^{r-1}$, though perhaps not with the expectation that we could reduce the number of generators to $r$.  Finally, we note that other sets of generators can be deduced from this one, in particular by using the method of $1$'s.  We leave this as an exercise for the enthusiastic reader.

\section{Discussion and Conclusions} \label{secConc}

In this paper we have attempted to give a complete account of our understanding regarding explicit, representation-theoretic presentations of the fusion rings and algebras associated to the Wess-Zumino-Witten models over the compact, connected, simply-connected (simple) Lie groups.  We have discussed presentations in terms of fusion potentials, and have provided complete proofs of the fact that there are explicitly known potentials which correctly describe the fusion \emph{algebras} of the models over $\func{\group{SU}}{r+1}$ and $\func{\group{Sp}}{2r}$.  These potentials appear to have been guessed in an educated manner.  We hope that our proofs will complement what has already appeared in the literature, and will be useful for subsequent studies.  We have also proven that the fusion algebras of the other groups \emph{cannot} be described by potentials analogous to those known, which explains why attempts to guess these potentials have not been successful.

We recalled that it is the fusion \emph{ring}, rather than the fusion algebra, which is of physical interest in applications.  Despite the fact that the fusion ring is torsion-free, we noted that a presentation for the fusion algebra need not give a presentation of the fusion ring.  To overcome this, we have stated and proved a fairly elementary result (\propref{propGrobner}) giving an explicit presentation (that is easily constructed) of the fusion ring in all cases.  We believe that this is the first time such a presentation has been formulated.  It is in terms of (linear combinations) of irreducible characters, and so should be regarded as representation-theoretic in the strongest possible sense.

These general presentations have one rather obvious disadvantage in that the number of characters appearing is quite large.  Whilst easy to write down, these presentations nevertheless contain quite a bit of complexity.  However, we have seen that it is sometimes possible to express the relevant characters in terms of simpler characters, and so reduce the number of characters that appear.  In particular, we have used the well-known determinantal identities for the characters of $\func{\group{SU}}{r+1}$ and $\func{\group{Sp}}{2r}$ to \emph{derive} the fusion potentials from first principles.  An important corollary to our results is then that these fusion potentials correctly describe the fusion \emph{rings} of the $\func{\group{SU}}{r+1}$ and $\func{\group{Sp}}{2r}$ models.

We then extended this result to the $\func{\group{Spin}}{2r+1}$ models.  The corresponding determinantal identities for the characters did not lead to as nice a simplification as before, in particular we did not end up with a potential description, but the result, \eqnref{eqnFusRingGenB}, is still relatively concise.  To the best of our knowledge, this is the first rigorous representation-theoretic presentation of the fusion ideal (over $\CC$ or $\ZZ$) for these \WZW models.  Nonetheless, this presentation is not as concise as we would like for the concrete applications we have in mind.  Certainly, for our motivating application to D-brane charge groups, our result allows us to write down an explicit form for this group\footnote{The charge group has the form $\ZZ_x^{2^{r-2}}$ \cite{BraTwi}, and we can determine $x$ to be the greatest common divisor of the integers obtained by evaluating the fusion ideal generators at the origin of the weight space.  With respect to \eqnref{eqnFusRingGenB}, this amounts to replacing the complete symmetric polynomials $\func{H_m}{q , 1 , q^{-1}}$ by $\binom{m + 2r}{2r}$ (and then finding the greatest common divisor).}.  However, we have been unable to substantially simplify this formula, so as to rigorously prove the result conjectured in \cite{BouDBr02}.  We have checked that this result is numerically consistent (to high level) with the generators presented here.

We expect that this result can also be extended to the $\func{\group{Spin}}{2r}$ models.  However, we have not done so for two reasons.  First, as mentioned in \appref{secJTD}, the derivation of the appropriate determinantal identities requires a slightly more general approach than what we have been using.  It follows that the methods we applied in analysing the $\func{\group{Spin}}{2r+1}$ case will require an analagous generalisation.  However, we believe that this generalisation should follow easily from the methods used in \cite{FulRep}.  Our second reason in that as with the $\func{\group{Spin}}{2r+1}$ case, we do not expect to get as simple a presentation as we would like.  We feel that the root of this is the observation that determinants are not particularly well-suited to computations when the Weyl group is not a symmetric group.  A far more elegant approach would be to generalise the algebra of determinants to the other Weyl groups, and then derive ``generalised determinantal identities'' for the Lie group characters in terms of Weyl-symmetric polynomials.  It would be very interesting to see if such an approach can be constructed (if it has not already been), and we envisage that it may lead to more satisfactory fusion ring presentations.  We hope to return to this in the future.

\section*{Acknowledgements}

PB is financially supported by the Australian Research Council, and DR would like to thank the Australian National University for a visiting fellowship during this project.  We would also like to thank Arzu Boysal, Volker Braun and Howard Schnitzer for helpful and stimulating correspondence.

\appendix

\section{Determinantal Identities of Jacobi-Trudy Type} \label{secJT}

In this section, some formulae are presented, expressing the irreducible characters of the classical groups in terms of determinants of matrices whose entries are relatively simple characters.  These formulae, which we will call \emph{Jacobi-Trudy identities} are well-known for the groups $\func{\group{SU}}{r+1}$, $\func{\group{Sp}}{2r}$, $\func{\group{SO}}{2r+1}$, and $\func{\group{O}}{2r}$, and may be found in \cite{WeyCla,FulRep}.  We are not aware of a reference for the corresponding formulae for the spinor representations of $\func{\group{Spin}}{2r+1}$ or $\func{\group{Spin}}{2r}$, nor for the tensor representations of the latter which are not restrictions of $\func{\group{O}}{2r}$ representations.  We therefore indicate how Jacobi-Trudy identities for these cases may be derived, following the ``transcendental'' method of Weyl.

The transcendental method relies on Weyl's character formula \cite{WeyCla}:
\begin{equation*}
\chi_{\lambda} = \frac{A_{\lambda + \rho}}{A_{\rho}} \qquad \text{where} \qquad A_{\lambda} = \sum_{w \in \group{W}} \det w \ e^{\func{w}{\lambda}},
\end{equation*}
and an identity of Cauchy \cite{FulRep}:
\begin{equation} \label{eqnCauchy1}
\Det{\frac{1}{1 - x_i y_j}}_{i,j = 1}^k = \frac{\Det{x_i^{k-j}}_{i,j = 1}^k \Det{y_j^{k-i}}_{i,j = 1}^k}{\displaystyle \prod_{i,j = 1}^k \brac{1 - x_i y_j}}.
\end{equation}
Here, $\Det{a_{ij}}_{i,j = 1}^k$ denotes the determinant of the $k \times k$ matrix with entries $a_{ij}$.  An alternative form of Cauchy's identity is obtained by replacing $y_j$ by $y_j^{-1}$ and multiplying through:
\begin{equation*}
\Det{\frac{1}{y_j - x_i}}_{i,j = 1}^k = \frac{\Det{x_i^{k-j}}_{i,j = 1}^k \Det{y_j^{i-1}}_{i,j = 1}^k}{\displaystyle \prod_{i,j = 1}^k \brac{y_j - x_i}}.
\end{equation*}
We will often apply this in the form
\begin{equation} \label{eqnCauchy2}
\Det{\frac{t_j}{\brac{1 - q_i t_j} \brac{1 - q_i^{-1} t_j}}}_{i,j = 1}^k = \frac{\Det{\brac{q_i + q_i^{-1}}^{k-j}}_{i,j = 1}^k \Det{\brac{t_j + t_j^{-1}}^{i-1} t_j^k}_{i,j = 1}^k}{\displaystyle \prod_{i,j = 1}^k \brac{1 - q_i t_j} \brac{1 - q_i^{-1} t_j}},
\end{equation}
obtained by putting $x_i = q_i + q_i^{-1}$ and $y_j = t_j + t_j^{-1}$.

\subsection{$\mathbf{\func{\group{SU}}{r+1}}$} \label{secJTA}

The Weyl group is $\group{S}_{r+1}$, acting as permutations on the weights $\eps_i$ of the defining representation.  We put $q_i = e^{\eps_i}$, so $q_1 \cdots q_{r+1} = 1$, and write $\lambda = \sum_{i=1}^r \lambda^i \eps_i$, with $\lambda^{r+1} = 0$ (in particular, $\rho^j = r+1-j$).  Then, $\lambda^1 \geqslant \lambda^2 \geqslant \ldots \geqslant \lambda^{r+1} = 0$ are all integers, and
\begin{equation*}
A_{\lambda} = \Det{q_i^{\lambda^j}}_{i,j = 1}^{r+1}.
\end{equation*}
We would like to emphasise that the $\lambda^j$ are to be distinguished from the Dynkin labels, which we denote by $\lambda_j$.

We form a generating function and apply Cauchy's identity, \eqnref{eqnCauchy1}:
\begin{align*}
\sum_{\lambda^1 , \ldots , \lambda^{r+1} = 0}^{\infty} A_{\lambda} t_1^{\lambda^1} \cdots t_{r+1}^{\lambda^{r+1}} &= \Det{\sum_{\lambda^j = 0}^{\infty} q_i^{\lambda^j} t_j^{\lambda^j}}_{i,j = 1}^{r+1} = \Det{\frac{1}{1 - q_i t_j}}_{i,j = 1}^{r+1} \\
&= \frac{\Det{q_i^{r+1-j}} \Det{t_j^{r+1-i}}}{\displaystyle \prod_{i,j} \brac{1 - q_i t_j}}.
\end{align*}
We recognise $A_{\rho}$ in the numerator, and expand the denominator in terms of complete symmetric polynomials $\func{H_m}{q}$ in the $q_i$.  We then get
\begin{equation*}
\sum_{\lambda^1 , \ldots , \lambda^{r+1} = 0}^{\infty} \frac{A_{\lambda}}{A_{\rho}} t_1^{\lambda^1} \cdots t_{r+1}^{\lambda^{r+1}} = \Det{t_j^{r+1-i}} \prod_j \sqbrac{\sum_{m_j \in \ZZ} \func{H_{m_j}}{q} t_j^{m_j}}.
\end{equation*}
Bringing the symmetric polynomials into the determinant, changing the summation variables so that the power of $t_j$ is $\lambda^j + \rho^j$, and then bringing the $t_j$ out of the determinant finally gives the original Jacobi-Trudy identity:
\begin{equation} \label{eqnJTA}
\chi_{\lambda} = \Det{\func{H_{\lambda^j + i - j}}{q}}_{i,j = 1}^{r+1}.
\end{equation}
Note that applying this formula to $\lambda = m \Lambda_1 = m \eps_1$ gives $\func{H_m}{q} = \chi_{m \Lambda_1}$.

\subsection{$\mathbf{\func{\group{Sp}}{2r}}$} \label{secJTC}

This time the Weyl group is $\group{S}_r \ltimes \ZZ_2^r$, acting on the weights $\pm \eps_i$ of the defining representation by permutation ($\group{S}_r$) and sign flips (each $\ZZ_2$ negates one of the $\eps_i$ whilst leaving the others invariant).  With $\lambda = \sum_i \lambda^i \eps_i$, so $\rho^j = r+1-j$, we find
\begin{equation*}
A_{\lambda} = \Det{q_i^{\lambda^j} - q_i^{-\lambda_j}}_{i,j = 1}^r.
\end{equation*}
Here, $\lambda^1 \geqslant \lambda^2 \geqslant \ldots \geqslant \lambda^r \geqslant 0$ are all integers.  What follows is very similar to \appref{secJTA}, so the details are left to the reader.  The generating function this time gives the left-hand-side of \eqnref{eqnCauchy2}, up to a product $\prod_i \brac{q_i - q_i^{-1}}$.  After applying the alternative form of Cauchy's identity, this product combines with the $q$-determinant so obtained to give $A_{\rho}$.  From there, the story is as before, and we find that
\begin{equation} \label{eqnJTC}
\chi_{\lambda} = 
\begin{vmatrix}
\func{H_{\lambda^j + 1 - j}}{q,q^{-1}} \\
\func{H_{\lambda^j + i - j}}{q,q^{-1}} + \func{H_{\lambda^j + 2 - i - j}}{q,q^{-1}}
\end{vmatrix}
_{i,j = 1}^r.
\end{equation}
In this equation, the top entry of the matrix should be understood to describe the elements of row $i=1$, and the bottom entry describes the rows $i>1$.  The complete symmetric functions are in the $q_i$ and their inverses.  Note that $\func{H_m}{q,q^{-1}} = \chi_{m \Lambda_1}$.

\subsection{$\mathbf{\func{\group{Spin}}{2r+1}}$} \label{secJTB}

The Weyl group is again $\group{S}_r \ltimes \ZZ_2^r$, acting on the \emph{non-zero} weights $\pm \eps_i$ of the defining representation as in the $\func{\group{Sp}}{2r}$ case.  Therefore, we again find that
\begin{equation*}
A_{\lambda} = \Det{q_i^{\lambda^j} - q_i^{-\lambda_j}}_{i,j = 1}^r,
\end{equation*}
where $\lambda = \sum_i \lambda^i \eps_i$, and $\lambda^1 \geqslant \lambda^2 \geqslant \ldots \geqslant \lambda^r \geqslant 0$.  In contrast to the $\func{\group{Sp}}{2r}$ case, the $\lambda^i$ can either be all integers (corresponding to a representation of $\func{\group{SO}}{2r+1}$, also called a tensor representation) or all half-integers (a spinor representation).  Indeed, $\rho^j = r + \frac{1}{2} - j$.

If we form a generating function with $\lambda^j$ integral, \eqnref{eqnCauchy2} gives
\begin{equation*}
\sum_{\lambda^1 , \ldots , \lambda^r = 0}^{\infty} A_{\lambda} t_1^{\lambda^1} \cdots t_r^{\lambda^r} = \prod_i \brac{q_i - q_i^{-1}} \cdot \frac{\Det{\brac{t_j + t_j^{-1}}^{i-1} t_j^r} \Det{\brac{q_i + q_i^{-1}}^{r-j}}}{\displaystyle \prod_{i,j} \brac{1 - q_i t_j} \brac{1 - q_i^{-1} t_j}}.
\end{equation*}
Recognising that $A_{\rho}$ factors as $\prod_i \brac{q_i^{1/2} - q_i^{-1/2}} \cdot \Det{\brac{q_i + q_i^{-1}}^{r-j}}$, and proceeding as usual gives
\begin{equation} \label{eqnJTBS1}
\chi_{\lambda} = \prod_i \brac{q_i^{1/2} + q_i^{-1/2}} \cdot
\begin{vmatrix}
\func{H_{\lambda^j + \frac{1}{2} - j}}{q,q^{-1}} \\
\func{H_{\lambda^j - \frac{1}{2} + i - j}}{q,q^{-1}} + \func{H_{\lambda^j + \frac{3}{2} - i - j}}{q,q^{-1}}
\end{vmatrix}
_{i,j = 1}^r.
\end{equation}
Note that because the $\rho^j$ are half-integers, this describes the characters of the \emph{spinor} representations.  Note also that $\chi_r \equiv \chi_{\Lambda_r} = \prod_i \brac{q_i^{1/2} + q_i^{-1/2}}$.  Finally, as the defining representation has a zero weight, it may be more convenient to express this result in terms of the complete symmetric polynomials in the $q_i$, their inverses, and $1$.  This gives the Jacobi-Trudy identity for the spinor representations of $\func{\group{Spin}}{2r+1}$:
\begin{equation} \label{eqnJTBS}
\chi_{\lambda} = \chi_r \Det{\func{H_{\lambda^j - \frac{1}{2} + i - j}}{q,1,q^{-1}} - \func{H_{\lambda^j + \frac{1}{2} - i - j}}{q,1,q^{-1}}}_{i,j = 1}^r.
\end{equation}

Forming the generating function with $\lambda^j$ half-integral then gives the $\func{\group{Spin}}{2r+1}$ Jacobi-Trudy identity for the tensor representations.  The manipulations are straightforward, and give
\begin{equation} \label{eqnJTBT}
\chi_{\lambda} = \Det{\func{H_{\lambda^j + i - j}}{q,1,q^{-1}} - \func{H_{\lambda^j - i - j}}{q,1,q^{-1}}}_{i,j = 1}^r.
\end{equation}
Note that $\chi_{m \Lambda_1} = \func{H_m}{q,1,q^{-1}} - \func{H_{m-2}}{q,1,q^{-1}}$, so $\func{H_m}{q,1,q^{-1}} = \chi_{m \Lambda_1} + \chi_{\brac{m-2} \Lambda_1} + \ldots$.

Finally, if we compare \eqnref{eqnJTBS1} with \eqnref{eqnJTC}, we find that we have established a strange relationship between the characters of the spinor representations of $\func{\group{Spin}}{2r+1}$ and those of $\func{\group{Sp}}{2r}$.  This is perhaps best written in the following form, where $\lambda$ labels a tensor representation:
\begin{equation*}
\chi_{\lambda + \Lambda_r}^{\func{\group{Spin}}{2r+1}} = \chi_{\Lambda_r}^{\func{\group{Spin}}{2r+1}} \chi_{\lambda}^{\func{\group{Sp}}{2r}}.
\end{equation*}
(Of course, this has to be interpreted appropriately.)  Evaluating the characters at $0$ to get the dimensions of the corresponding representations gives an identity of \cite{GabCha}.  Interestingly, it is claimed there that this identity cannot hold at the level of characters.

\subsection{$\mathbf{\func{\group{Spin}}{2r}}$} \label{secJTD}

The Weyl group is $\group{S}_r \ltimes \ZZ_2^{r-1}$, acting on the weights $\pm \eps_i$ of the defining representation as in the $\func{\group{Sp}}{2r}$ case, except that the $\ZZ_2^{r-1}$ factor corresponds to transformations where an \emph{even} number of the $\eps_i$ are negated and the rest are left invariant.  Therefore, 
\begin{equation} \label{eqnWeylAltD}
2 A_{\lambda} = \Det{q_i^{\lambda^j} + q_i^{-\lambda_j}}_{i,j = 1}^r + \Det{q_i^{\lambda^j} - q_i^{-\lambda_j}}_{i,j = 1}^r,
\end{equation}
where $\lambda = \sum_i \lambda^i \eps_i$, and $\lambda^1 \geqslant \lambda^2 \geqslant \ldots \geqslant \lambda^{r-1} \geqslant \abs{\lambda^r}$.  As in the previous case, we have tensor representations ($\lambda^i \in \ZZ$) and spinor representations ($\lambda^i \in \ZZ + \frac{1}{2}$).  A non-trivial Dynkin diagram symmetry (for $r > 4$ this is the only such symmetry) acts via $\lambda^r \rightarrow - \lambda^r$, so representations with $\lambda^r = 0$ will be referred to as symmetric\footnote{For $r$ odd, this symmetry is conjugation, so symmetric coincides with self-conjugate.  However, for $r$ even, the conjugation automorphism is trivial.}.  Symmetric representations correspond to representations of $\func{\group{O}}{2r}$, and it is clear that for these representations, the second term in the above formula for $A_{\lambda}$ vanishes.  Note that $\rho^j = r - j$ defines a symmetric (tensor) representation:
\begin{equation*}
A_{\rho} = \frac{1}{2} \Det{q_i^{r - j} + q_i^{-\brac{r - j}}}_{i,j = 1}^r = \Det{\brac{q_i + q_i^{-1}}^{r - j}}_{i,j = 1}^r.
\end{equation*}

Since $\rho$ is tensor, forming a generating function with each $\lambda^j$ half-integral and positive gives an identity for spinor representations.  The derivation of this identity should by now be an easy exercise for the reader.  It is:
\begin{multline*}
2 \chi_{\lambda} = \prod_i \brac{q_i^{1/2} + q_i^{-1/2}} \cdot \Det{\func{H_{\lambda^j - \frac{1}{2} + i - j}}{q,q^{-1}} - \func{H_{\lambda^j + \frac{1}{2} - i - j}}{q,q^{-1}}}_{i,j = 1}^r \\
+ \prod_i \brac{q_i^{1/2} - q_i^{-1/2}} \cdot \Det{\func{H_{\lambda^j - \frac{1}{2} + i - j}}{q,q^{-1}} + \func{H_{\lambda^j + \frac{1}{2} - i - j}}{q,q^{-1}}}_{i,j = 1}^r.
\end{multline*}
Setting all $\lambda^i = \frac{1}{2}$ gives $2 \chi_{r-1} = \prod_i \brac{q_i^{1/2} + q_i^{-1/2}} + \prod_i \brac{q_i^{1/2} - q_i^{-1/2}}$.  As we assumed $\lambda^i \geqslant \frac{1}{2}$ when computing the generating function, this formula cannot be applied to $\chi_r$ directly.  Instead, it is determined from $\chi_{r-1}$ by applying the Dynkin symmetry $q_r \rightarrow q_r^{-1}$ (this symmetry has the effect of changing the sign of the second term in the above equation).  Thus, $2 \chi_r = \prod_i \brac{q_i^{1/2} + q_i^{-1/2}} - \prod_i \brac{q_i^{1/2} - q_i^{-1/2}}$, leading to the $\func{\group{Spin}}{2r}$ Jacobi-Trudy identity for spinor representations:
\begin{multline} \label{eqnJTDS}
\chi_{\lambda} = \frac{1}{2} \brac{\chi_{r-1} + \chi_r} \Det{\func{H_{\abs{\lambda^j} - \frac{1}{2} + i - j}}{q,q^{-1}} - \func{H_{\abs{\lambda^j} + \frac{1}{2} - i - j}}{q,q^{-1}}}_{i,j = 1}^r \\
\pm \frac{1}{2} \brac{\chi_{r-1} - \chi_r} \Det{\func{H_{\abs{\lambda^j} - \frac{1}{2} + i - j}}{q,q^{-1}} + \func{H_{\abs{\lambda^j} + \frac{1}{2} - i - j}}{q,q^{-1}}}_{i,j = 1}^r.
\end{multline}
The $\pm$ appearing here reflects the sign of $\lambda^r$.  Of course, the absolute values appearing in $\abs{\lambda^j}$ are only necessary for $j = r$.

The corresponding derivation for tensor representations is somewhat unique in that Weyl's transcendental method does not seem to be directly applicable to the first term in \eqnref{eqnWeylAltD}.  Instead, we have to resort to the algebraic method (see \cite{FulRep}).  Weyl's method has no problem with the second term, so this hybrid gives the $\func{\group{Spin}}{2r}$ Jacobi-Trudy identity for tensor representations:
\begin{equation} \label{eqnJTDT}
\chi_{\lambda} = 
\begin{cases}
\Det{\func{H_{\lambda^j + i - j}}{q,q^{-1}} - \func{H_{\lambda^j - i - j}}{q,q^{-1}}}_{i,j = 1}^r & \text{if $\lambda^r = 0$,} \\
\frac{1}{2} \Det{\func{H_{\abs{\lambda^j} + i - j}}{q,q^{-1}} - \func{H_{\abs{\lambda^j} - i - j}}{q,q^{-1}}}_{i,j = 1}^r & \\
\mspace{20mu} \pm \frac{1}{2} \brac{\chi_{r-1}^2 - \chi_r^2} 
\begin{vmatrix}
\func{H_{\abs{\lambda^j} - j}}{q,q^{-1}} \\
\func{H_{\abs{\lambda^j} - 1 + i - j}}{q,q^{-1}} + \func{H_{\abs{\lambda^j} + 1 - i - j}}{q,q^{-1}}
\end{vmatrix}
_{i,j = 1}^r & \text{if $\lambda^r \neq 0$.}
\end{cases}
\end{equation}
Again, the $\pm$ reflects the sign of $\lambda^r$ and correlates with the application of the Dynkin symmetry $q_r \rightarrow q_r^{-1}$.  Note that $\chi_{m \Lambda_1} = \func{H_m}{q,q^{-1}} - \func{H_{m-2}}{q,q^{-1}}$.  We also note that $\chi_{r-1}^2 - \chi_r^2 = \chi_{2 \Lambda_{r-1}} - \chi_{2 \Lambda_r}$.

\subsection{Further Remarks} \label{secJTOther}

Comparing these $\func{\group{Spin}}{2r}$ identities to those derived for the other groups, we note two novelties.  One is the fact that two determinants are generally required, and the second is that explicit factors of $\frac{1}{2}$ appear (in spite of the fact that the right hand side must be a polynomial in the fundamental characters with integral coefficients).  These novelties are direct consequences of the form of \eqnref{eqnWeylAltD}, which itself reflects the increasing complexity of the Weyl group of $\func{\group{Spin}}{2r}$, as compared to the cases already treated.  Roughly speaking, the Weyl group is sufficiently ``non-symmetric'' (where ``symmetric'' refers to the symmetric group) that the use of determinants in Weyl's transcendental method, in particular applying Cauchy's identity (\eqnDref{eqnCauchy1}{eqnCauchy2}), leads to annoyingly complicated Jacobi-Trudy identities.

The Weyl groups of the exceptional groups are even less ``symmetric'', and so we expect that the above methods used to derive Jacobi-Trudy identities will be next to useless in these cases.  Indeed, the simplest exceptional group $\group{G}_2$ has the dihedral group of order $12$ for its Weyl group:  $\group{W} = \group{D}_{12} = \ZZ_2 \ltimes \group{S}_3$.  Na\"{\i}vely proceeding with Weyl's transcendental method leads to the evaluation of an unpleasant quotient.  Forcing the evaluation with the aid of a computer suggests that the corresponding Jacobi-Trudy identity may require as many as \emph{sixty} determinants!

The appropriate course of action seems therefore clear.  Rather than try to force determinants unnaturally upon a Weyl group in order to apply Cauchy's identity, we should instead try to generalise Cauchy's identity in such a way that it applies to Weyl's alternants $A_{\lambda} = \sum_{w \in \group{W}} \det w \ e^{\func{w}{\lambda}}$ directly.  We are not aware of any such generalisation, but given the magic of Weyl groups, we would not be surprised if such a generalisation could be found.  We speculate that such a finding may lead to simple and useful identities of Jacobi-Trudy type for all simple Lie groups.

\end{document}